\documentclass[aps,prd,twocolumn,groupedaddress,nofootinbib]{revtex4}

\pdfoutput=1

\usepackage[autostyle, english = british]{csquotes}
\MakeOuterQuote{"}

\usepackage[english]{babel}

\usepackage{graphicx}
\usepackage{adjustbox}
\usepackage{subfigure}
\usepackage{capt-of}
\usepackage{bbm}
\usepackage{amsmath,amssymb,amsfonts,tensor}
\usepackage{xcolor}

\usepackage{enumitem}
\setlist[itemize,1]{topsep=2mm,parsep=-1mm,leftmargin=1.5cm}

\usepackage{physics} 
\usepackage[percent]{overpic} 
\usepackage{pdfpages}

\usepackage{hyperref}
\usepackage{cleveref}

\hypersetup{colorlinks,citecolor=blue,linkcolor=blue,urlcolor=blue,pdfencoding=auto, psdextra}

\newcommand{\be}{\begin{equation}}
\newcommand{\ee}{\end{equation}}
\newcommand*{\diff}{\mathop{}\!\mathrm{d}}

\newcommand{\D}{{{\cal D}_\gamma^{(j)}}}
\newcommand{\Df}{{\cal D}_\gamma^{(j_{\mathrm{f}})}}
\newcommand{\DF}{{\cal D}_\gamma^{(j_{\mathrm{F}})}}

\begin{document}

\title{The End of a Black Hole's Evaporation -- Part II}

\author{Farshid Soltani${}^{a,b}$, Carlo Rovelli${}^{c,d,e}$ and Pierre Martin-Dussaud${}^{f}$}

\affiliation{\vspace{.25cm}${}^a$ Department of Physics and Astronomy, University of Western Ontario, London, ON N6A 3K7, Canada}
\affiliation{${}^b$ Dipartimento di Fisica, Sapienza Universit\`a di Roma, P.le Aldo Moro 5, I-00185 Roma, EU}
\affiliation{${}^c$ AMU Universit\'e, Universit\'e de Toulon, CNRS, CPT, F-13288 Marseille, EU}
\affiliation{${}^d$ Perimeter Institute, 31 Caroline Street N, Waterloo ON, N2L2Y5, Canada} 
\affiliation{${}^e$ The Rotman Institute of Philosophy, 1151 Richmond St.~N London  N6A5B7, Canada}
\affiliation{${}^f$ Institute for Gravitation and the Cosmos, The Pennsylvania State University, University Park, Pennsylvania 16802, USA
\vspace{.15cm}}

\date{\small\today}


\begin{abstract}
\noindent 
We derive an explicit expression for the transition amplitude from black to white hole horizon at the end of Hawking evaporation using covariant loop quantum gravity. 
\end{abstract}

\maketitle 

\section{Introduction}

\noindent
The existence of a classical solution of the Einstein field equations describing a spacetime in which the exterior of a past black hole and a future white hole are connected \cite{Haggard2015a} indicates that the end of the evaporation of a black hole can result in a quantum tunnelling from a trapped to an anti-trapped region.
The black to white hole transition is receiving increasing attention in the literature \cite{Modesto2004,Hossenfelder2010a,GambiniPullin2014a,Olmedo2017,Malafarina2017,Ashtekar2018b,Ashtekar2018d,Clements2019a,Bodendorfer2019a,Volovik2021,BenAchour2020,Munch2020,Kelly2021}. In \cite{Dambrosio2021}, following the scenario developed in \cite{Rovelli2014,Haggard2014,Haggard2016,DeLorenzo2016,Christodoulou2016,Rovelli2018a,DAmbrosio2018a,Christodoulou2018d,Bianchi2018c,Martin-Dussaud2019}, we have introduced a technique to study this transition. Here we use this technique to derive an explicit expression for the corresponding transition amplitude.

This transition amplitude is formally given by a path integral over four-geometries performed on a spacetime region $B$ bounded by a three-dimensional surface $\Sigma$ (with specified intrinsic and extrinsic geometry). The three-dimensional surface $\Sigma$ encloses the transition region. Its specification and its geometry have been computed in \cite{Dambrosio2021}. The formal path integral is approximated and concretely defined by the spinfoam amplitude \cite{book:Rovelli_Vidotto_CLQG} associated to a discretization of $B$. While the discretization of $\Sigma$ was defined in \cite{Dambrosio2021}, here we construct the full discretization of $B$ and compute the corresponding transition amplitude. 

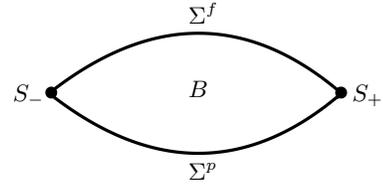
\begin{figure}
    \begin{center}
    \begin{overpic}[width = 0.6 \columnwidth]{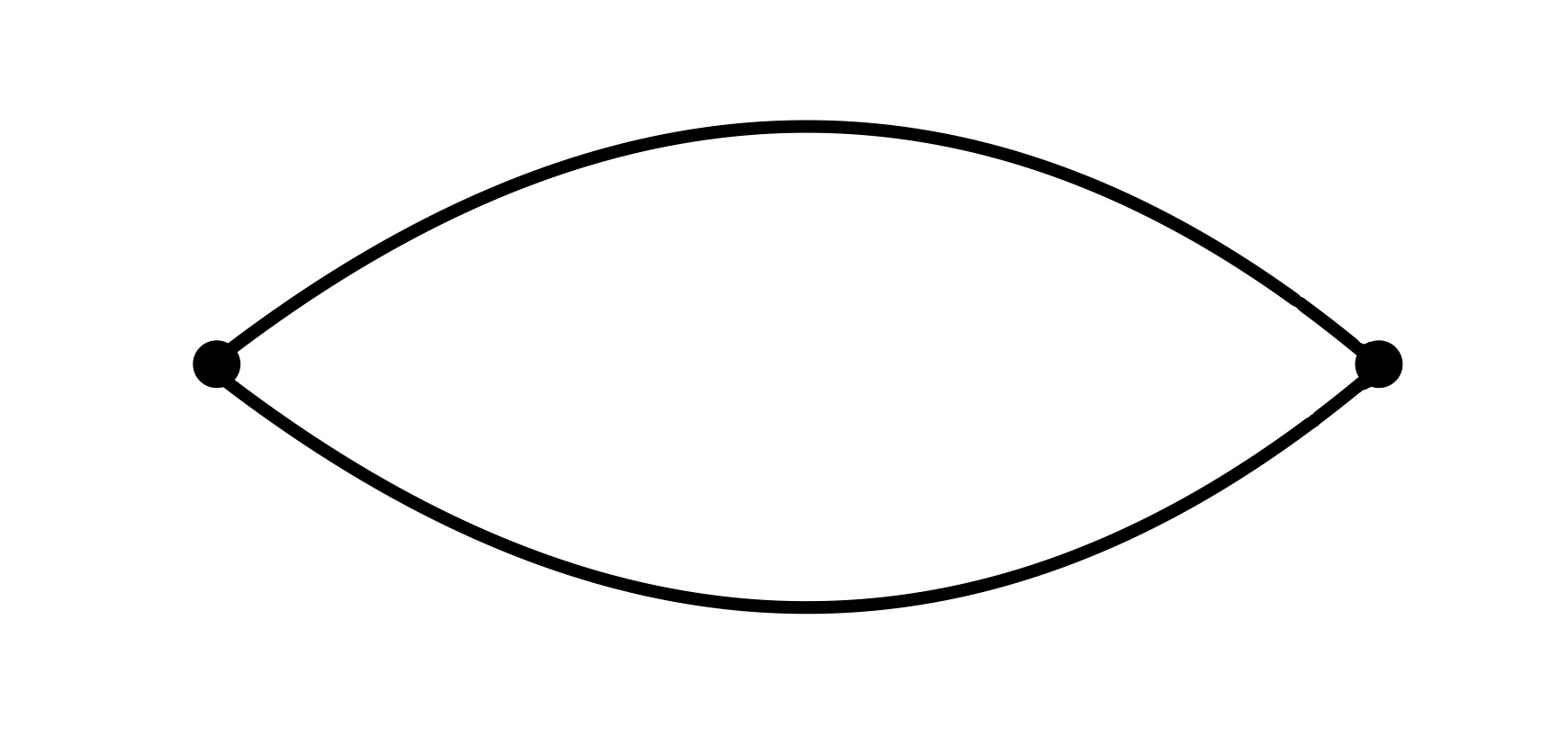}
    \put (49,2) {$\Sigma^p$}
    \put (49,42) {$\Sigma^f$}
    \put (91,22) {$S_+$}
    \put (4,22) {$S_-$}
    \put (49,23) {$B$}
    \end{overpic}
    \end{center}
    \caption{The region $B$ in the time-radius space: each point of the diagram is a two sphere.}
    \label{fig:region_B}
\end{figure}

The complexity of the calculation is given by the topology of $B$, which is the product of a two-sphere and a disk (see \cref{fig:region_B}). 
The disk is the product of a finite time interval and a finite space (radial) interval and it is delimited by an exterior two-sphere $S_+$ that surrounds the horizon and by an interior two-sphere $S_-$ surrounded by the horizon (sitting on the bounce radius of the transition of the internal geometry of the black hole). The two two-spheres $S_+$ and $S_-$ split $\Sigma$ into a past component $\Sigma^p$ and a future component $\Sigma^f$.

While $\Sigma$ was discretized by means of a three-dimensional triangulation, here we discretize $B$ in terms of a two-complex $\mathcal{C}$ dual to a cellular complex which is not a four-dimensional triangulation. This choice has the advantage of providing a relatively simple discretization that respects the symmetries of the problem. 

\Cref{section:discretization} is devoted to the construction of the two-complex $\mathcal{C}$. The corresponding transition amplitude is computed in \cref{section:amplitude}. \Cref{section:symmetry} offers a simplification of the expression based on the symmetries of the two-complex. \Cref{cs} gives the amplitude in terms of coherent states. The duals of $\Gamma$ and $\mathcal{C}$, namely the triangulation of $\Sigma$ and the cellular decomposition of $B$, are discussed in \cref{section:appendix_1}. \Cref{section:appendix_2} offers a graphical representation of $\Gamma$ and $\mathcal{C}$. \Cref{section:appendix_3} recalls the basic formulas of covariant loop quantum gravity.

\section{Discretization of \texorpdfstring{$B$}{B}}
\label{section:discretization}

\noindent
In this section we give the combinatorial definition of the two-complex $\mathcal{C}$ as a set of vertices, edges and faces with their boundary relations. To help the geometric visualization, a graphical representation of the two-complex $\mathcal{C}$ and its boundary $\Gamma$ is provided in \cref{section:appendix_2}: we advise the reader to consult it. 

If $\mathrm{N}_1$ and $\mathrm{N}_2$ are nodes, we write $\mathrm{L}=(\mathrm{N}_1,\mathrm{N}_2)$ to denote the oriented link  with source $\mathrm{N}_1$ and  target $\mathrm{N}_2$. We denote $\mathrm{L}^{-1}\equiv(\mathrm{N}_2,\mathrm{N}_1)$ the same link but with opposite orientation. For the vertices and edges of the two-complex of the spinfoam (which form a graph), we use an analogous notation. We denote the vertices as $\mathrm{v}$; the internal edges (bounded by two vertices) as $\mathrm{e}$; the external edges (bounded by one node) as $\mathrm{E}$. Similarly, we denote the internal faces (bounded by internal edges only) as $\mathrm{f}$; the external faces (with one link in the boundary) as $\mathrm{F}$. We write $\mathrm{f}=(\mathrm{e}_{1},\,\dots\,,\mathrm{e}_{n})$ to denote the oriented face $\mathrm{f}$ bounded by these edges. The orientation of the face is given by the sequence of edges. These are written oriented accordingly to the orientation the face induces on them. 

Let $a,b,c,d$ be indices taking values in the set $\{1,2,3,4\}$, $t$ be an index taking values in the set $\{p,f\}$ (for past and future) and $\epsilon$ be an index taking values in the set $\{-,+\}$ (for interior and exterior). In all the expressions with several indices $a,b,\dots$ these are assumed to be all different, that is $a\ne b$ and so on. Unless two indices are separated by a comma, the order of the indices is not relevant and exchanging the indices results in the same element. If two indices are separated by a comma, exchanging the indices results in a different element. 

The graph $\Gamma$ dual to the three-dimensional triangulation of $\Sigma$ constructed in \cite{Dambrosio2021} is defined by
\begin{itemize}
\item[Nodes\  \  ]  $\mathrm{N}^{t\epsilon}_a$ and  $\mathrm{N}^{t}_{ab}$;\vspace{1mm}
\item[Links\ \ \   ] $\mathrm{L}^{\epsilon}_{a}=
(\mathrm{N}^{p \epsilon}_a , \mathrm{N}^{f \epsilon}_a)$,  \vspace{1mm}\\
$\mathrm{L}^{t+}_{a,b}=(\mathrm{N}^{t}_{ab},\mathrm{N}^{t+}_a)$, \vspace{.5mm}\\
$\mathrm{L}^{t-}_{a,b}=(\mathrm{N}^{t}_{cd},\mathrm{N}^{t-}_a)$.
\end{itemize}
(See figure \ref{fig:Gamma_2}  of \cref{section:appendix_2}.)

The two-complex $\cal C$, whose boundary $\partial \!\: \cal C$ is given by $\Gamma$, is defined by:
\begin{itemize}
\item[Vertices\ ] $\mathrm{v}^\epsilon_{a}$ and $\mathrm{v}_{ab}$; \vspace{1mm}
\item[Edges\ \ ]  $\mathrm{E}^{t\epsilon}_a=(\mathrm{v}^\epsilon_{a},\mathrm{N}^{t\epsilon}_a)$, \vspace{.5mm}\\ 
$\mathrm{E}^t_{ab}=(\mathrm{v}_{ab},\mathrm{N}^{t}_{ab})$,  \vspace{.5mm}\\
$\mathrm{e}^+_{a,b}=(\mathrm{v}^+_a,\mathrm{v}_{ab})$,\vspace{.25mm}\\
$\mathrm{e}^-_{a,b}=(\mathrm{v}^-_a,\mathrm{v}_{cd})$; \vspace{1mm}
\item[Faces\ \ ]   $\mathrm{F}^{\epsilon}_{a}=\big(\mathrm{L}^{\epsilon}_{a},(\mathrm{E}^{f\epsilon}_a)^{-1},\mathrm{E}^{p\epsilon}_a\big)$, \vspace{1mm}\\ $\mathrm{F}^{t+}_{a,b}=\big(\mathrm{L}^{t+}_{a,b},
(\mathrm{E}^{t+}_a)^{-1},\mathrm{e}^{+}_{a,b},
\mathrm{E}^t_{ab}\big)$, 
\vspace{1mm}\\ 
$\mathrm{F}^{t-}_{a,b}=\big(\mathrm{L}^{t-}_{a,b},
(\mathrm{E}^{t-}_a)^{-1},\mathrm{e}^{-}_{a,b},
\mathrm{E}^t_{cd}\big)$, 
\vspace{1mm}\\ 
$\mathrm{f}_{a,b} \overset{c<d}{=} \big( \mathrm{e}^+_{a,c},(\mathrm{e}^-_{b,d})^{-1},\mathrm{e}^-_{b,c},(\mathrm{e}^+_{a,d})^{-1} \big) $. 
\end{itemize}
This construction defines the orientation of every element of the two-complex. 

The geometry of $B$ is invariant under both rotations and the time reversal transformation that swaps $\Sigma^p$ and $\Sigma^f$. The discretization reduces the rotational symmetry to a discrete tetrahedral symmetry, realized by an even permutation of the indices $a,b,c,d$. The time reversal symmetry is realized by the swap of the indices $p$ and $f$. 

There is also a combinatorial symmetry defined by the exchange of the exterior and the interior, namely by the exchange of $S_+$ and $S_-$. This is realized in the two-complex by the invariance under the swap of the indices $+$ and $-$. This is however not a symmetry of the geometry we want to study, as $S_+$ and $S_-$ have a different geometry: $S_+$  is larger. 

\section{Transition amplitude}
\label{section:amplitude}

\noindent
Following \cref{section:appendix_3}, we assign group elements to the edges and the links of the two-complex:
\begin{itemize}
\item[] $\mathrm{L}^{\epsilon}_{a} \longleftrightarrow h^{\epsilon}_{a} \in \mathrm{SU}(2)\, ;$
\item[] $\mathrm{L}^{t\epsilon}_{a,b}\longleftrightarrow h^{t\epsilon}_{a,b} \in \mathrm{SU}(2)\, ;$
\item[] $\mathrm{E}^{t\epsilon}_a \longleftrightarrow g^{t \epsilon}_{a} \in \mathrm{SL}(2,\mathbb{C})\, ;$
\item[] $\mathrm{E}^t_{ab} \longleftrightarrow g^{t}_{ab} \in \mathrm{SL}(2,\mathbb{C})\, ;$
\item[]$ \mathrm{e}^{\epsilon}_{a,b} \longleftrightarrow g^{\epsilon}_{a\rightarrow b}\, , \, g^{\epsilon}_{a\leftarrow b}\in \mathrm{SL}(2,\mathbb{C})\, .$
\end{itemize}
The group element $g^{\epsilon}_{a\rightarrow b}$ is assigned to the oriented half-edge of $\mathrm{e}^{\epsilon}_{a,b}$ having source in the source of $\mathrm{e}^{\epsilon}_{a,b}$ and target in the middle of $\mathrm{e}^{\epsilon}_{a,b}$. The group element $g^{\epsilon}_{a\leftarrow b}$ is assigned to the oriented half-edge of $\mathrm{e}^{\epsilon}_{a,b}$ having source in the target of $\mathrm{e}^{\epsilon}_{a,b}$ and target in the middle of $\mathrm{e}^{\epsilon}_{a,b}$.
 
It is then straightforward to compute the covariant loop quantum gravity transition amplitude for the black-to-white hole transition by using the expressions reported in \cref{section:appendix_3} applied to the two-complex defined above. The two-complex amplitude $W_{\mathcal{C}}$ expressed in terms of face amplitudes is
\be
\begin{split}
 W_{\mathcal{C}} \big(h^\epsilon_a, & h^{t\epsilon}_{a,b}\big) 
 =  \int_{\mathrm{SL}(2,\mathbb{C})}  
\dd g^{p\epsilon}_a \: \dd g^p_{ab}\  \dd g^\epsilon_{a\leftrightarrow b}  \\
    & \times 
    \prod_{a\epsilon}   A_{a}^\epsilon(h^{\epsilon}_{a},g^{t\epsilon}_a )   
   \prod_{ab} A_{a,b}(g^+_{a\leftrightarrow c},g^-_{b\leftrightarrow c} )   \\ 
 & \times \prod_{t ab} A^{t+}_{a,b}(h^{t+}_{a,b},g^{t+}_a, g^t_{ab},g^+_{a\leftrightarrow b}) \\
 & \times \prod_{t ab} A^{t-}_{a,b}(h^{t-}_{a,b},g^{t-}_a, g^t_{cd},g^-_{a\leftrightarrow b})\, .
 \label{equation:transition_amplitude_faces}
\end{split}
\ee
To regularize the expression in \eqref{equation:transition_amplitude_faces} we have dropped the integration over one $\mathrm{SL}(2,\mathbb{C})$ element per vertex. We have chosen to drop the integrations over the $g^{f\epsilon}_{a}$ and $g^{f}_{ab}$ variables. The integral is independent from these.

To write the face amplitudes we use the following notation. We introduce the $(2j+1)\times (2j+1)$ matrix $\D(g)$, $g\in \mathrm{SL}(2,\mathbb{C})$, with matrix elements  
\be
(\D(g))_{mn}\equiv {\cal D}^{(\gamma j, j)}_{jm\: jn} (g)\, ,
\label{calD}
\ee
where ${\cal D}^{(p, k)}_{jm\: j'n}$ are the matrix elements of the $(p,k)$ unitary representation of  the principal series of $\mathrm{SL}(2,\mathbb{C})$ in the canonical basis that diagonalize the operators $L^2$ and $L_z$ of the $\mathrm{SU}(2)$ subgroup \cite{book:Rovelli_Vidotto_CLQG}.  The matrix $\D(g)$ should not be confused with the $\mathrm{SU}(2)$ Wigner matrix $D^{(j)}(h)$ with matrix elements $D^{(j)}_{mn}(h)$, $h\in \mathrm{SU}(2)$. Using these, we have
\be
\begin{split}
A^{\epsilon}_{a}( h^{\epsilon}_{ a} ,g^{t\epsilon}_a )=\displaystyle\sum_{j} d_j \ 
& \Tr\Big[\D\big( (g^{f\epsilon}_{a})^{-1} \: g^{p\epsilon}_{a} \big) \\
 & \times
D^{(j)}\big( h^{\epsilon}_{ a} \big)\Big]\, ,
\label{equation_face_amplitude_ea}
\end{split}
\ee
\be
\begin{split}
A_{a,b} & (g^+_{a\leftrightarrow c},g^-_{b\leftrightarrow c} ) \overset{c<d}{=} \displaystyle\sum_{j}      d_j 
\Tr\Big[\D\big( (g^{+}_{a\rightarrow d})^{-1} \: g^{+}_{a\rightarrow c} \big)\\
  \times \: &
\D\big( (g^{+}_{a\leftarrow c})^{-1} \: g^{-}_{b\leftarrow d} \big)\ \D\big( (g^{-}_{b\rightarrow d})^{-1} \: g^{-}_{b\rightarrow c} \big) \\  \times \: &
\D\big( (g^{-}_{b\leftarrow c})^{-1} \: g^{+}_{a\leftarrow d} \big)\Big]
\, , 
\end{split}
\label{equation_face_amplitude_ab}
\ee
\begin{eqnarray}
A^{t+}_{a,b} (h^{t+}_{a,b},g^{t+}_a, g^t_{ab},g^+_{a\leftrightarrow b} )
= \displaystyle\sum_{j} d_j\,  \Tr\Big[
\D\big( (g^{t+}_{a})^{-1} \: g^{+}_{a\rightarrow b} \big)  
\nonumber \\ 
\times   \D\big( (g^{+}_{a\leftarrow b})^{-1} \:\: g^{t}_{ab}   \big) 
D^{(j)}\big( h^{t+}_{a,b}     \big) \Big] \, , \quad \quad \quad
\label{equation:face_amplitude_t+ab}
\end{eqnarray}
\begin{eqnarray}
A^{t-}_{a,b} (h^{t-}_{a,b},g^{t-}_a, g^t_{cd},g^-_{a\leftrightarrow b} )
= \displaystyle\sum_{j} d_j\,  \Tr\Big[
\D\big( (g^{t-}_{a})^{-1} \: g^{-}_{a\rightarrow b} \big)  
\nonumber \\ 
\times   \D\big( (g^{-}_{a\leftarrow b})^{-1} \:\: g^{t}_{cd}   \big) 
D^{(j)}\big( h^{t-}_{a,b}     \big) \Big] \, .  \quad  \quad \quad
\label{equation:face_amplitude_t-ab}
\end{eqnarray}

The transition amplitude for the black-to-white hole transition is then given by 
\be
\begin{split}
\bra{W_{\mathcal{C}}}\ket{\psi_{\mathrm{BW}}}= &\int_{\mathrm{SU}(2)} 
\diff  h^\epsilon_a \diff h^{t\epsilon}_{a,b} \\
\times &
W_{\mathcal{C}} ( h^\epsilon_a,h^{t\epsilon}_{a,b})\;
\psi_{\mathrm{BW}} ( h^\epsilon_a,h^{t\epsilon}_{a,b})\: ,
\label{equation:appendix_LQG_state_amplitude2}
\end{split}
\ee
where $\ket{\psi_{\mathrm{BW}}}\in \mathcal{H}_{\Gamma}$ is the extrinsic coherent state peaked on the classical boundary geometry of the black-to-white hole transition that was constructed in \cite{Dambrosio2021}. 

For numerical calculations \cite{Dona2018a,Dona2019a,Dona2020b,gozzini2021highperformance} and to use asymptotic techniques \cite{article:Speziale_2017_Boosting_nj_symbols,Dona2020c,Dona2020d} it is more convenient to use the amplitude written in terms of vertex amplitudes. 

\begin{figure}
  \includegraphics[keepaspectratio=true,scale=0.4]{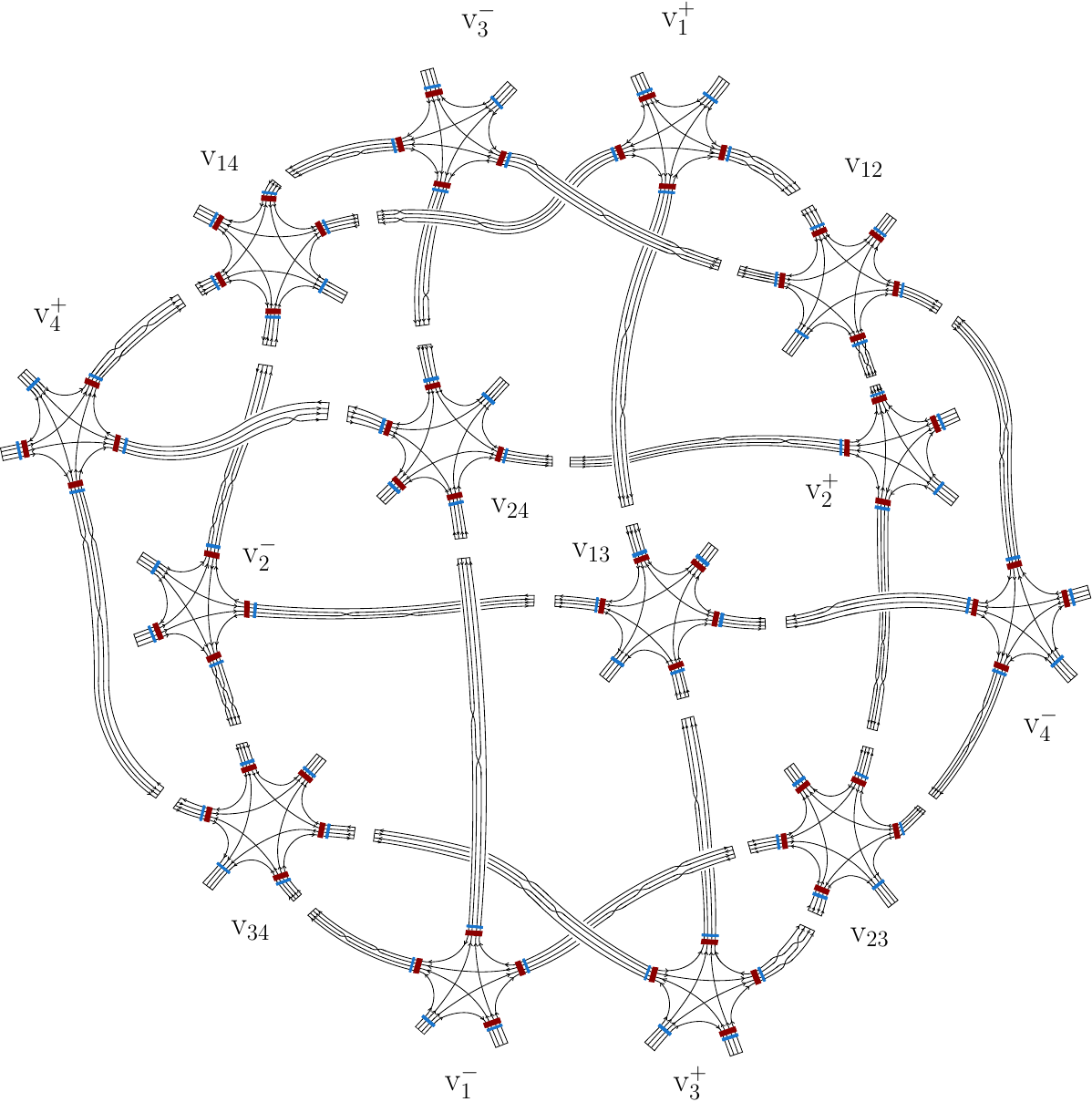}
\captionof{figure}{Graphical representation of the product of the local vertex amplitudes entering the two-complex spin-intertwiner transition amplitude in \cref{equation:coherent_transition_amplitude2}.}
\label{fig:graphical_representation_intertwiner}
\end{figure}

Following the notation in \cref{section:appendix_3}, we assign a spin (an $\mathrm{SU}(2)$ irreducible representation) to every face in the two-complex and an intertwiner, namely a basis element in the space (each edge of the two-complex belongs to four faces)
\be
H_{j_1\cdots j_4}= \mathrm{Inv}_{\mathrm{SU}(2)} \big[H_{j_1}\otimes\cdots\otimes H_{j_4}\big],
\ee 
to every edge of the two-complex:
\begin{itemize}
\item[] $\mathrm{F}^{\epsilon}_{a} \longleftrightarrow J^{\epsilon}_{a}\in \mathbb{N}/2 \, ;$
\item[] $\mathrm{F}^{t\epsilon}_{a,b}\longleftrightarrow J^{t\epsilon}_{a,b} \in \mathbb{N}/2 \, ;$
\item[] $\mathrm{f}_{a,b} \longleftrightarrow j_{a,b}\in \mathbb{N}/2 \, ;$
\item[] $\mathrm{E}^{t\epsilon}_{a} \longleftrightarrow 
{I}^{t\epsilon}_{a} \in H_{J^\epsilon_aJ^{t\epsilon}_{a,b}} \,;$
\item[]$ \mathrm{E}^{t}_{ab} \longleftrightarrow {I}^{t}_{ab} \in H_{J^{t+}_{a,b}J^{t-}_{c,d}} \,;$

\item[] $\mathrm{e}^{+}_{a,b} \longleftrightarrow {i}^{+}_{a,b} \in H_{j_{a,c}J^{t+}_{a,b}} \,;$
\item[] $\mathrm{e}^{-}_{a,b} \longleftrightarrow {i}^{-}_{a,b} \in H_{j_{c,a}J^{t-}_{a,b}} \,;$
\end{itemize}
Using this we can write the two-complex transition amplitude in the following form:
\be
\begin{split}
W_{\mathcal C}& \big(h^\epsilon_a, h^{t\epsilon}_{a,b}  \big)  =  
\sum_{J^{\epsilon}_{a} \,J^{t\epsilon}_{a,b}} \sum_{{I}^{t\epsilon}_{a} \,  {I}^{t}_{ab}} 
\Big[ \prod_{\epsilon a} d_{J^{\epsilon}_{a}} 
\prod_{t \epsilon a b}  d_{J^{t\epsilon}_{a,b}} \Big] 
\\\times & 
\bra{\bigotimes_{t\epsilon a} I^{t\epsilon}_a  \bigotimes_{ta,b>a}I^{t}_{ab} 
}\ket{\bigotimes_{\epsilon a} D^{(J^{\epsilon}_{a})} ( h^{\epsilon}_{a})
\bigotimes_{t\epsilon ab} D^{(J^{t\epsilon}_{a,b})} ( h^{t\epsilon}_{a,b})}_\Gamma
  \\
\times & \ \ \ 
W_{\mathcal C} \big( J^{\epsilon}_{a} \,J^{t\epsilon}_{a,b},
 {I}^{t\epsilon}_{a} ,  {I}^{t}_{ab}
    \big) \, ,
\end{split}
\label{equation:coherent_transition_amplitude6}
\ee
where the bra-ket notation indicates the index contraction dictated by $\Gamma$ \cite{book:Rovelli_Vidotto_CLQG}.
The transition amplitude in the spin-intertwiner basis reads
\be
\begin{split}
W_{\mathcal C} & \big(J^{\epsilon}_{a}  \,J^{t\epsilon}_{a,b},
 {I}^{t\epsilon}_{a} ,  {I}^{t}_{ab})   =
  \sum_{j_{a,b}}\sum_{i^\epsilon_{a,b}} 
 \prod_{ a b}  d_{j_{a,b}} 
 \\ 
 \times &  \ \prod_{\epsilon a}
 A_{\mathrm{v}^{\epsilon}_a}\big(j_{a,b}, J^{\epsilon}_{a},J^{t\epsilon}_{a,b},
{I}^{t\epsilon}_{a} , i^\epsilon_{a,b})
 \\
 \times &
\prod_{a}\prod_{b>a}    A_{\mathrm{v}_{ab}}
\big(j_{a,c},j_{b,c},J^{t+}_{a,b}, J^{t-}_{c,d}, {I}^{t}_{ab},i^+_{a,b},i^-_{c,d} \big) \: .
\end{split}
\label{equation:coherent_transition_amplitude2}
\ee
The graphical representation of the product of the vertex amplitudes (whose analytical expression can be easily read from \cref{equation:appendix_LQG_vertex_amplitude_intertwiner}) written using the graphical representation of \cite{Perez:2012wv,article:Speziale_2017_Boosting_nj_symbols} can be found in \cref{fig:graphical_representation_intertwiner}.

Notice the relative simplicity of the expression in \cref{equation:coherent_transition_amplitude2}. There are only two kind of local vertex amplitudes (see also \cref{fig:graphical_representation_intertwiner}): the amplitude associated to the eight five-valent vertices $\mathrm{v}_a^\epsilon$ and the amplitude associated to the six six-valent vertices $\mathrm{v}_{ab}$.  

For example, the amplitude of the five-valent vertex $\mathrm{v}_1^+$ depends on the following quantities. Two "vertical" (see \cref{section:appendix_2}) boundary intertwiners, the past one ${I}^{p+}_{1} $ and the future one $ {I}^{f+}_{1}$, and three "horizontal" internal intertwiners $i^+_{1,2}, i^+_{1,3}, i^+_{1,4}$. One "vertical" spin $J_1^+$, three "vertical" past spins $J^{p}_{1,2},J^{p}_{1,3},J^{p}_{1,4}$, three "vertical" future spins $J^{f}_{1,2},J^{f}_{1,3},J^{f}_{1,4}$ and finally three spins $j_{1,2}, j_{1,3}, j_{1,4}$ of the "horizontal" internal faces. 

Similarly, the amplitude of the six-valent vertex $\mathrm{v}_{12}$ depends on the following quantities. Two "vertical" boundary intertwiners ${I}^{p}_{12} $ and $ {I}^{f}_{12}$ and four "horizontal" internal intertwiners $i^+_{1,2}, i^+_{2,1}, i^-_{3,4},i^-_{4,3}$. Four "vertical" past spins $J^{p+}_{1,2},J^{p+}_{2,1},J^{p-}_{3,4},J^{p-}_{4,3}$, four "vertical" future spins $J^{f+}_{1,2},J^{f+}_{2,1},J^{f-}_{3,4},J^{f-}_{4,3}$ and four spins $j_{1,3},j_{1,4},j_{2,3},j_{2,4}$ of the "horizontal" internal faces.

\section{Using the symmetry}
\label{section:symmetry}

\noindent
Focusing on the transition amplitude in the spin-intertwiner basis in \cref{equation:coherent_transition_amplitude2}, a simplification of the model can be obtained by restricting the computation to spin foam configurations whose colouring respect the geometric symmetries of the problem at hand. Namely, we can simplify the amplitude by performing a symmetry reduction of the model. 

In this scenario the boundary spins are fixed to just four independent ones,
\be
 J^\epsilon_a= \tilde J_\epsilon \;\quad\text{and}\quad\; J^{t\epsilon}_{a,b}=  J_\epsilon \, ,
 \ee
and the boundary intertwiners are fixed to just three independent ones,
\be
 I^{t\epsilon}_{a}=  I_\epsilon \;\quad\text{and}\quad\;
 I^{t}_{a,b}=  I \, .
 \ee
Furthermore, simple geometry shows that the symmetry of the tetrahedra dual to the exterior and the interior nodes $\mathrm{N}^{t\epsilon}_a$ (see \cref{section:appendix_1}) implies that their geometry is fully determined by the area of their faces \cite{Dambrosio2021}. Hence, the two boundary intertwiners $I_\epsilon$ are actually uniquely determined by the spins: $I_\epsilon=I_\epsilon(\tilde J_\epsilon, J_\epsilon)$. Finally, the symmetric truncation of the model is obtained by restricting to symmetric spin foam configurations, that is
\be
j_{a,b}=j   \;\quad\text{and}\quad\;
i^{\epsilon}_{a,b}= i_{\epsilon}.
\ee
The transition amplitude in the spin-intertwiner basis of the symmetry reduced model is then given by
\be
\begin{split}
W_{\mathcal C} \big(J_{\epsilon},  \,\tilde J_\epsilon, {I})   =
  \sum_{j,i_\epsilon} & \:\: d^{12}_{j} 
   \:\:A^6 \big(j,J_{\epsilon},
 {I},i_\epsilon ) \: 
 \\ \times & 
  \prod_\epsilon 
 A^4_{\epsilon}\big(j, J_{\epsilon}, \tilde J_\epsilon, i_\epsilon)\,, 
\end{split}
\label{equation:coherent_transition_amplitude3}
\ee
\begin{figure}
\includegraphics[scale=0.05]{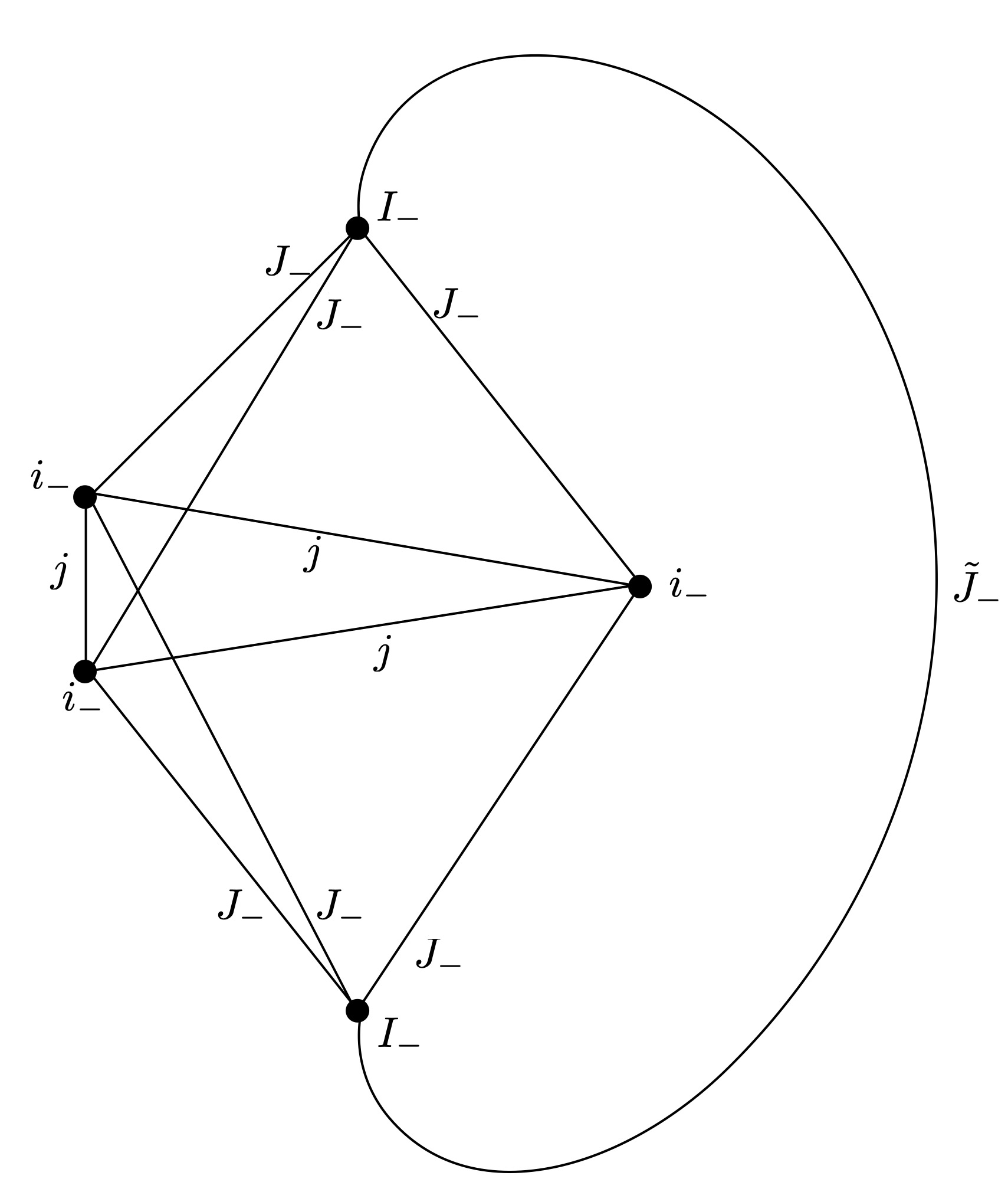}\ \ \ \ 
\includegraphics[scale=0.13]{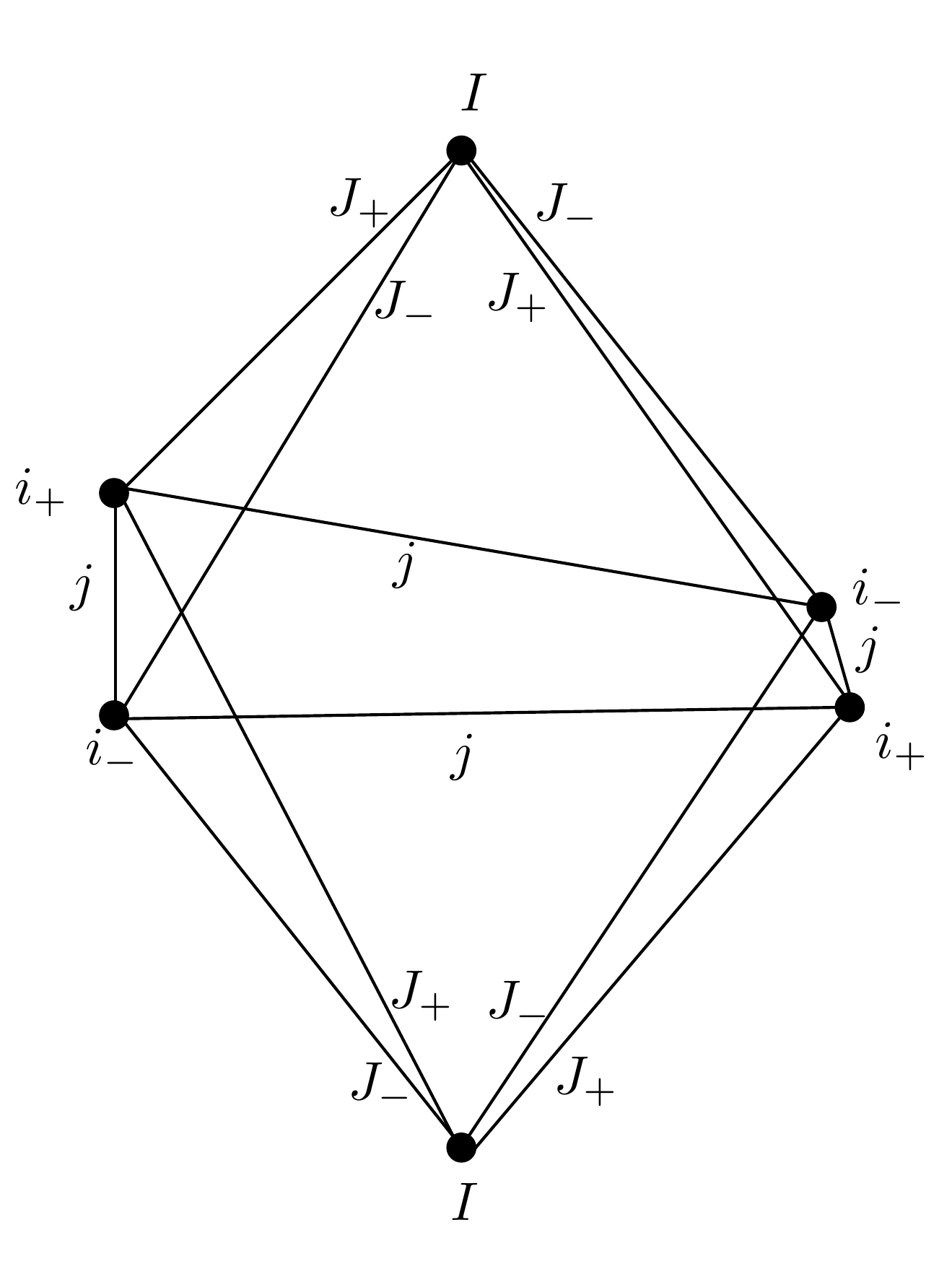}
\captionof{figure}{The boundary graph \cite{book:Rovelli_Vidotto_CLQG} of the interior five-valent vertices (left) and of the six-valent vertices (right) in the symmetry reduced model.}
\label{v5}        
\end{figure}
where $A \big(j,J_{\epsilon}, {I},i_\epsilon )$ is the local vertex amplitude associated to every $\mathrm{v}_{ab}$ vertex and $A_{\epsilon}\big(j, J_{\epsilon}, \tilde J_\epsilon, i_\epsilon)$ is the local vertex amplitude associated to every $\mathrm{v}^{\epsilon}_{a}$ vertex. This is a huge simplification with respect to the transition amplitude in \cref{equation:coherent_transition_amplitude2}.

The explicit dependence of the vertex amplitudes on the spins and the intertwiners can be read from \cref{v5} or from \cref{fig:graphical_representation_intertwiner}.

\section{Coherent state basis}\label{cs}

\noindent 
Finally, we give also the expression of the transition amplitude in the over-complete basis on the intertwiner spaces formed by the coherent intertwiners. This form of the amplitude is useful to study the intrinsic quantum geometry of the cellular complex dual to the two-complex.

Following again the notation in \cref{section:appendix_3}, we assign a normal to every couple (edge, face) in the two-complex:
\begin{itemize}
\item[] $\big(\mathrm{E}^{t\epsilon}_{a}, \mathrm{F}^{\epsilon}_{a}\big) \longleftrightarrow 
\vec{N}^{t\epsilon}_{a} \in S^2 \,;$
\item[]$ \big(\mathrm{E}^{t\epsilon}_{a}, \mathrm{F}^{t\epsilon}_{a,b}\big) 
\longleftrightarrow \vec{N}^{t\epsilon}_{a,b} \in S^2 \,;$
\item[]$ \big(\mathrm{E}^{t}_{ab}, \mathrm{F}^{t+}_{a,b}\big) \longleftrightarrow \vec{N}^{t+}_{ab,a} \in S^2 \,;$
\item[]$ \big(\mathrm{E}^{t}_{ab}, \mathrm{F}^{t-}_{c,d}\big) \longleftrightarrow \vec{N}^{t-}_{ab,c} \in S^2 \,;$
\item[] $\big(\mathrm{e}^{\epsilon}_{a,b}, \mathrm{F}^{t\epsilon}_{a,b}\big) \longleftrightarrow \vec{n}^{t\epsilon}_{a,b} \in S^2 \,;$
\item[] $\big(\mathrm{e}^{+}_{a,b}, \mathrm{f}_{a,c}\big) \longleftrightarrow \vec{n}^{+}_{a,b,c} \in S^2 \,;$
\item[] $\big(\mathrm{e}^{-}_{a,b}, \mathrm{f}_{c,a}\big) \longleftrightarrow \vec{n}^{-}_{a,b,c} \in S^2 \,.$
\end{itemize}

The change of basis gives
\be
\begin{split}
W_{\mathcal C}  \big(h^\epsilon_a , &  h^{t\epsilon}_{a,b}   \big)  =  \sum_{J^{\epsilon}_{a} \,J^{t\epsilon}_{a,b} } 
\Big[ \prod_{\epsilon a} d_{J^{\epsilon}_{a}} 
\prod_{t \epsilon a b}  d_{J^{t\epsilon}_{a,b}} \Big] \\
\times &   \int_{S^2} 
\Big( \diff^2 \vec{N}^{t\epsilon}_{a} \: \frac{d_{J^{\epsilon}_{a}}}{4\pi}  \Big)
\Big( \diff^2 \vec{N}^{t+}_{ab,a} \:\frac{d_{J^{t+}_{a,b}}}{4\pi}  \Big)
\\ \times & \ \ \ \ 
\Big( \diff^2 \vec{N}^{t-}_{ab,c} \:\frac{d_{J^{t-}_{c,d}}}{4\pi}  \Big)
\Big( \diff^2 \vec{N}^{t\epsilon}_{a,b} \:\frac{d_{J^{t \epsilon}_{a,b}}}{4\pi}  \Big)
\\\times & \prod_{\epsilon a}
\bra{J^{\epsilon}_{a},\vec{N}^{p\epsilon}_{a}}
D^{(J^{\epsilon}_{a})} ( h^{\epsilon}_{a})
\ket {J^{\epsilon}_{a},\vec{N}^{f\epsilon}_{a}}     \\
\times & \prod_{t a b}
\bra{J^{t+}_{a,b},\vec{N}^{t+}_{ab,a}}
D^{(J^{t+}_{a,b})} ( h^{t+}_{a,b})
\ket {J^{t+}_{a,b},\vec{N}^{t+}_{a,b}}     \\
\times & \prod_{t a b}
\bra{J^{t-}_{a,b},\vec{N}^{t-}_{cd,a}}
D^{(J^{t-}_{a,b})} ( h^{t-}_{a,b})
\ket {J^{t-}_{a,b},\vec{N}^{t-}_{a,b}}     \\
\times & 
W_{\mathcal C} \big( J^{\epsilon}_{a} \,J^{t\epsilon}_{a,b},
\vec{N}^{t\epsilon}_{a}, 
 \vec{N}^{t+}_{ab,a}, 
 \vec{N}^{t-}_{ab,c},  \vec{N}^{t\epsilon}_{a,b} 
    \big), 
\end{split}
\label{equation:coherent_transition_amplitude4}
\ee
where the transition amplitude in the coherent basis reads
\be
\begin{split}
W_{\mathcal C} & \big(J^{\epsilon}_{a} , J^{t\epsilon}_{a,b},
\vec{N}^{t\epsilon}_{a}, 
 \vec{N}^{t+}_{ab,a}, 
 \vec{N}^{t-}_{ab,c},  \vec{N}^{t\epsilon}_{a,b})   =
  \sum_{j_{a,b} } 
 \prod_{ a b}  d_{j_{a,b}} 
 \\  \times &
\int_{S^2} 
\Big( \diff^2 \vec{n}^{t\epsilon}_{a,b} \:\frac{d_{J^{t\epsilon}_{a,b}}}{4\pi}  \Big)
\Big( \diff^2 \vec{n}^{+}_{a,b,c} \:\frac{d_{j_{a,c}}}{4\pi}  \Big)
\Big( \diff^2 \vec{n}^{-}_{a,b,c} \:\frac{d_{j_{c,a}}}{4\pi}  \Big)\\
\times &   \prod_{\epsilon a}
 A_{\mathrm{v}^{\epsilon}_a}\big(j_{a,b}, J^{\epsilon}_{a},J^{t\epsilon}_{a,b},
 \vec{n}^{t\epsilon}_{a,b},  \vec{n}^{\epsilon}_{a,b,c}, \vec{N}^{t\epsilon}_{a}, \vec{N}^{t\epsilon}_{a,b}) \\
 \times &
\prod_{a} \prod_{b>a}  A_{\mathrm{v}_{ab}}\big(j_{\mathrm{v}_{ab}},J_{\mathrm{v}_{ab}},\vec{n}_{\mathrm{v}_{ab}}, \vec{N}_{\mathrm{v}_{ab}} \big)  
\end{split}
\label{equation:coherent_transition_amplitude5}
\ee
and the labels $j_{\mathrm{v}_{ab}},J_{\mathrm{v}_{ab}},\vec{n}_{\mathrm{v}_{ab}}, \vec{N}_{\mathrm{v}_{ab}}$ stand for the following sets of variables:
\begin{subequations}\label{eq:def_setvariables}
\begin{align}
&j_{\mathrm{v}_{ab}}= \{ j_{a,c},j_{b,c}\}, \\
&J_{\mathrm{v}_{ab}}= \{ J^{t+}_{a,b}, J^{t-}_{c,d} \}, \\
&\vec{n}_{\mathrm{v}_{ab}}= \{ \vec{n}^{t+}_{a,b}, \vec{n}^{t-}_{c,d},  \vec{n}^{+}_{a,b,c}, \vec{n}^{-}_{c,d,a} \}, \\
&\vec{N}_{\mathrm{v}_{ab}}= \{ \vec{N}^{t+}_{ab,a}, \vec{N}^{t-}_{ab,c} \} .
\end{align}
\end{subequations}
The analytical expression of the coherent vertex amplitudes in \cref{equation:coherent_transition_amplitude5} is given in \cref{equation:appendix_LQG_vertex_amplitude}. The contraction pattern between the coherent states and the $\D$ matrices in the coherent vertex amplitudes exactly matches the contraction pattern between the intertwiners and the $\D$ matrices in the spin-intertwiner vertex amplitudes (it is just a change of basis) and it can therefore be read directly from \cref{fig:graphical_representation_intertwiner}


\section{Concluding remarks}

\noindent 
Starting from the results obtained in \cite{Dambrosio2021}, we have constructed a two-complex that discretizes the quantum region in which the black hole horizon tunnels from a trapping to an anti-trapping horizon and we have explicitly computed the transition amplitude associated to the phenomenon using covariant loop quantum gravity.

The two-complex defined in \cref{section:discretization} is complicated and the corresponding transition amplitude computed in \cref{section:amplitude} may be difficult to studied analytically. However, the high degree of symmetry of the two-complex should help in the numerical exploration of the transition amplitude. A numerical analysis of the transition amplitude is in progress and it will be reported elsewhere.

Notably, the set of internal faces of the two-complex forms a \textit{bubble}, that is they form together a surface without boundary with the topology of a two-sphere. The presence of a bubble in a two-complex may in principle lead to a divergence in the corresponding transition amplitude~\cite{book:Rovelli_Vidotto_CLQG}. 
A preliminary counting of integration variables and constraints among those appearing in the large spin limit of \eqref{equation:coherent_transition_amplitude5} appears to suggest that the amplitude is exponentially suppressed for large spins, and therefore convergent. A detailed analysis will be reported elsewhere.

\acknowledgments{\noindent This work was made possible through the support of the  FQXi  Grant  FQXi-RFP-1818 and of the ID\# 61466 grant from the John Templeton Foundation, as part of the The Quantum Information Structure of Spacetime (QISS) Project (\href{qiss.fr}{qiss.fr}). }

\vspace{1cm}

\appendix

\section{The cellular decomposition of \texorpdfstring{$B$}{B}}
\label{section:appendix_1}

\noindent
The triangulation of the boundary $\Sigma$ of $B$ was constructed in \cite{Dambrosio2021} and it can be summarized as follows.

The surface $\Sigma$ is formed by a past component $\Sigma^p$ and a future component $\Sigma^f$, joined inside the black hole at the two-sphere $S_-$ and outside the black hole at the two-sphere $S_+$ (see \cref{fig:region_B}). A simple triangulation of it is obtained by placing 4 (equidistant) points $p_a^-$ on $S_-$ and 4 (equidistant) points $p_a^+$ on $S_+$. 

\begin{figure}
    \centering
    \begin{overpic}[width = 0.9\columnwidth]{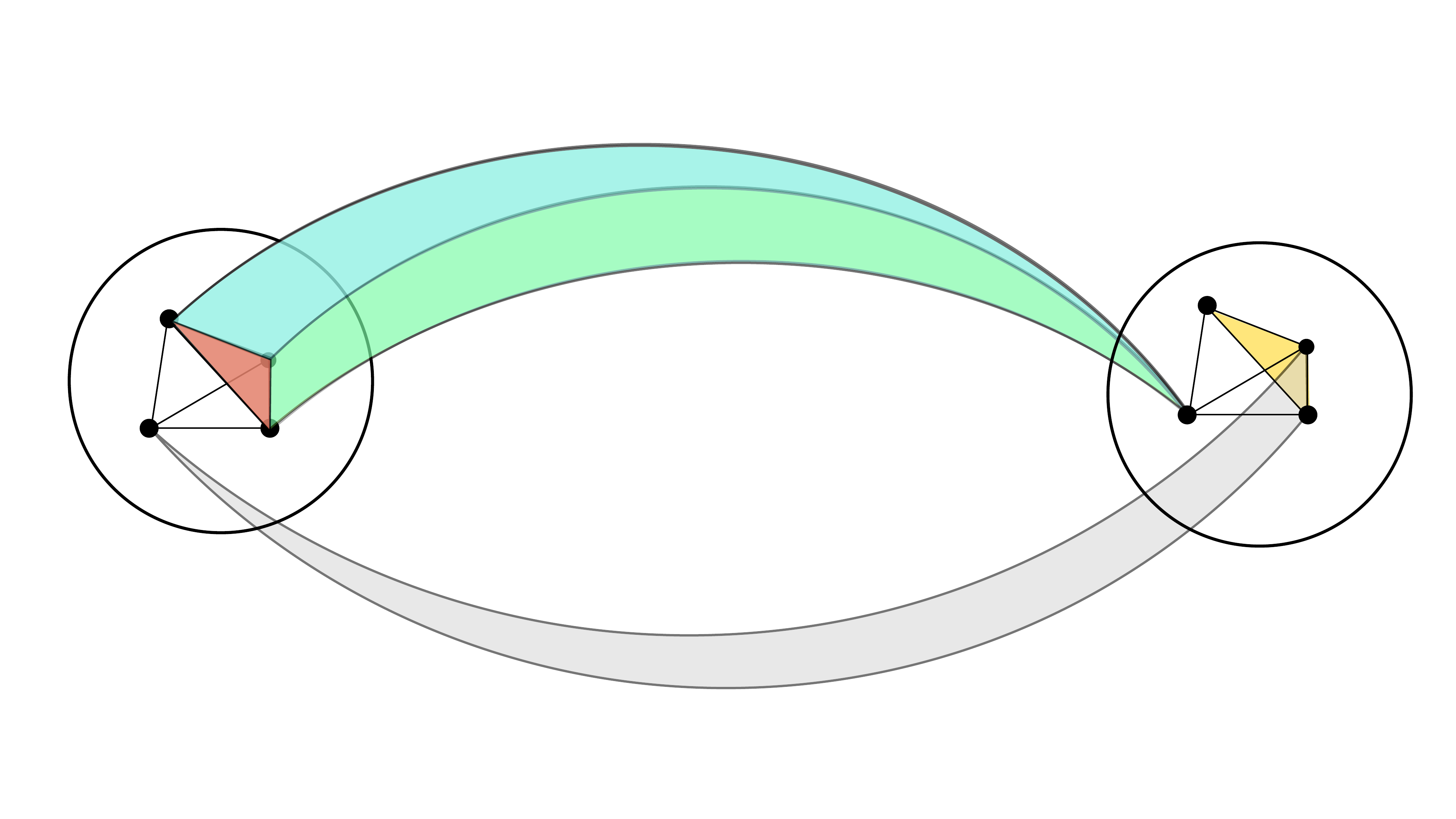}
    \put (49,49) {$\Sigma^f$}
    \put (49,4) {$\Sigma^p$}
    \put (90,15) {$S_+$}
    \put (8.5,15) {$S_-$}
    \put (77,25) {\scriptsize $p_1^+$}
    \put (90,25) {\scriptsize $p_2^+$}
    \put (91,32) {\scriptsize $p_4^+$}
    \put (79,35) {\scriptsize $p_3^+$}
    \put (7,24) {\scriptsize $p_1^-$}
    \put (20,25) {\scriptsize $p_2^-$}
    \put (7.5,35) {\scriptsize $p_3^-$}
    \put (35,39) {$N_1^{f-}$}
    \put (60,12) {$L^{p+}_{1,3}$}
    \put (85,35) {$L_1^+$}
    \put (5,30) {$s_{13}^-$}
    \end{overpic}
    \vspace{-2mm}
    \begin{overpic}[width = 0.9\columnwidth]{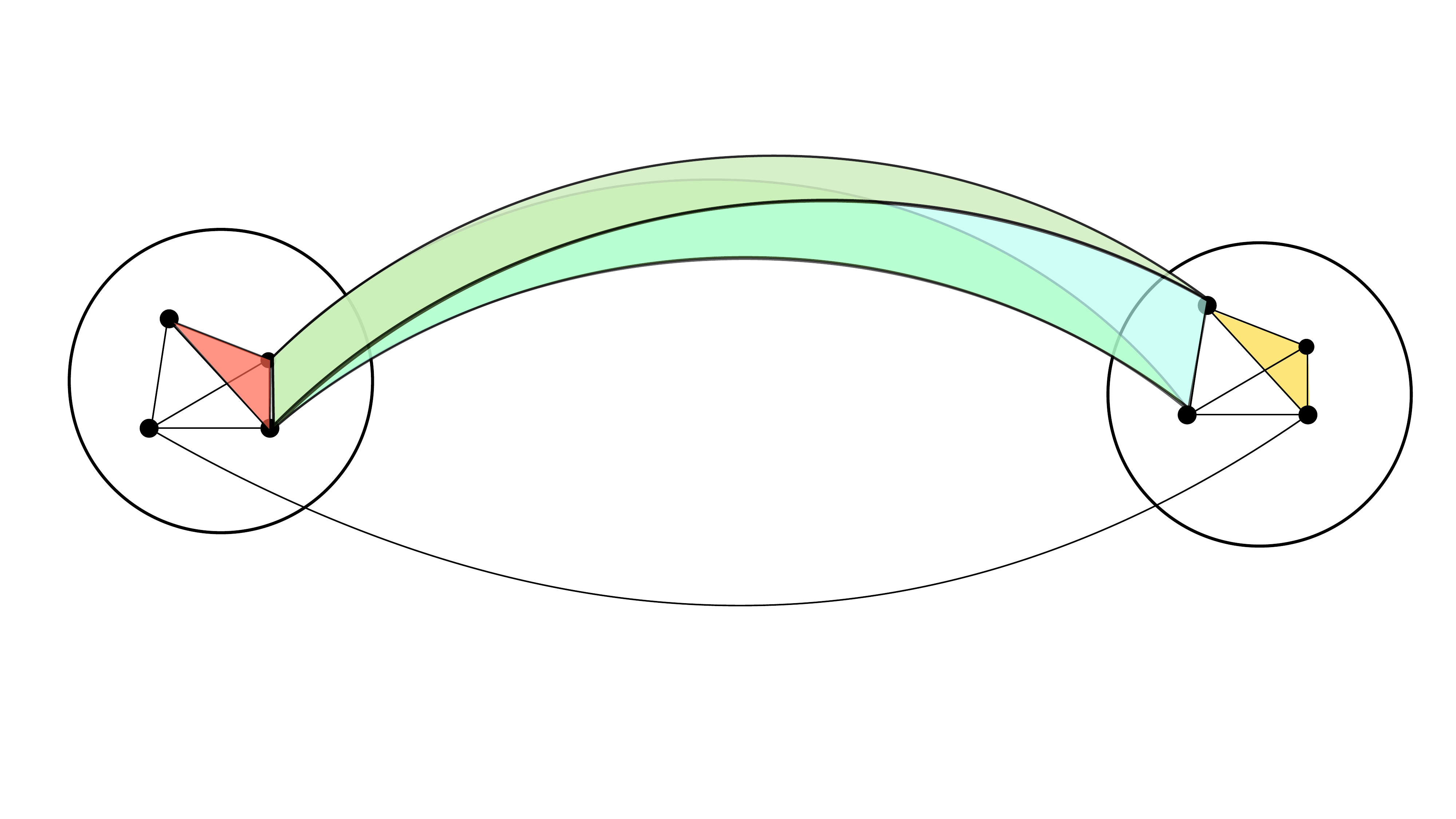}
    \put (40,47) {$\Sigma^f$}
    \put (40,9.5) {$\Sigma^p$}
    \put (90,15) {$S_+$}
    \put (8.5,15) {$S_-$}
    \put (77,25) {\scriptsize $p_1^+$}
    \put (90,25) {\scriptsize $p_2^+$}
    \put (91,32) {\scriptsize $p_4^+$}
    \put (7,24) {\scriptsize $p_1^-$}
    \put (20,25) {\scriptsize $p_2^-$}
    \put (7.5,35) {\scriptsize $p_3^-$}
    \put (16,34) {\scriptsize $p_4^-$}
    \put (55,41) {$N_{24}^{f}$}
    \put (60,12) {$s_{1,2}^p$}
    \put (85,35) {$L_1^+$}
    \put (5,30) {$s_{13}^-$}
    \end{overpic}
    \caption{Elements of the triangulation of $\Sigma$ seen as embedded objects in the four-dimensional spacetime.}
    \label{fig:triangulation_Sigma_fancy}
\end{figure}
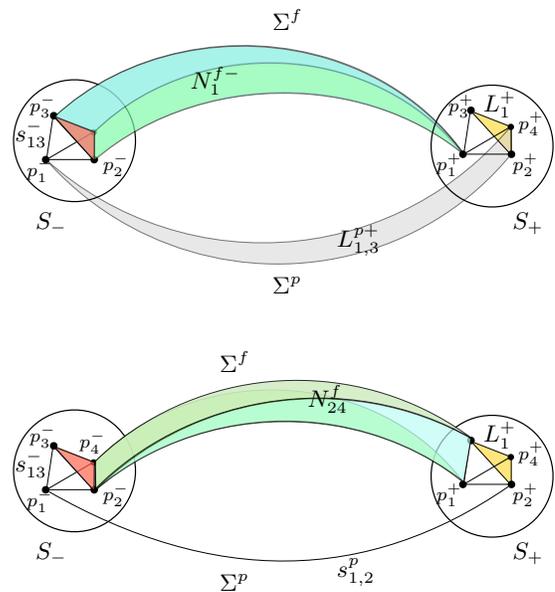

\begin{figure}
\centering
  \subfigure[][]{%
  \label{fig:triangulation_sigma_t_plain}%
  \includegraphics[scale=0.15]{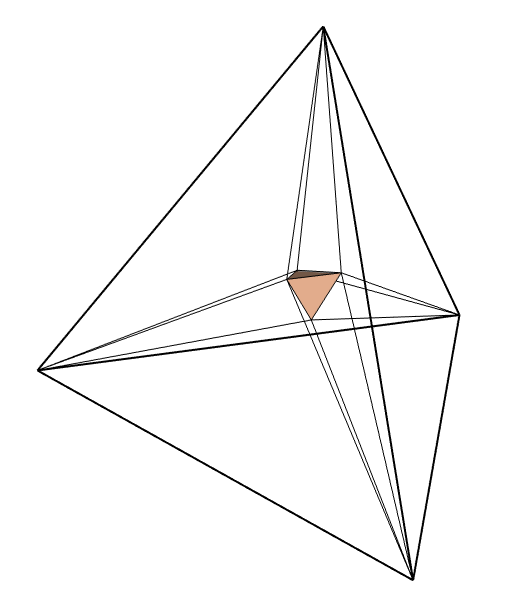}
  }%
  \hspace{22pt}
  \subfigure[][]{%
  \label{fig:triangulation_sigma_t_T-}%
  \includegraphics[scale=0.17]{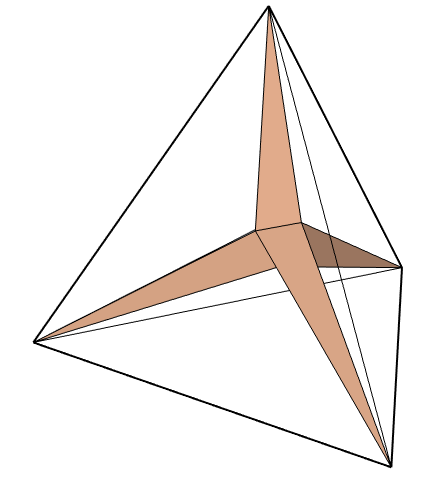}
  }%
  \hspace{29pt}
  \subfigure[][]{%
  \label{fig:triangulation_sigma_t_T+}%
  \includegraphics[scale=0.17]{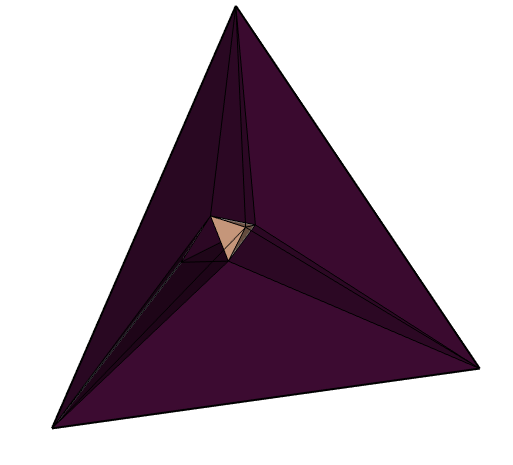}
  }
    \hspace{25pt}\subfigure[][]{%
  \label{fig:triangulation_sigma_t_Tab_alto}%
  \includegraphics[scale=0.15]{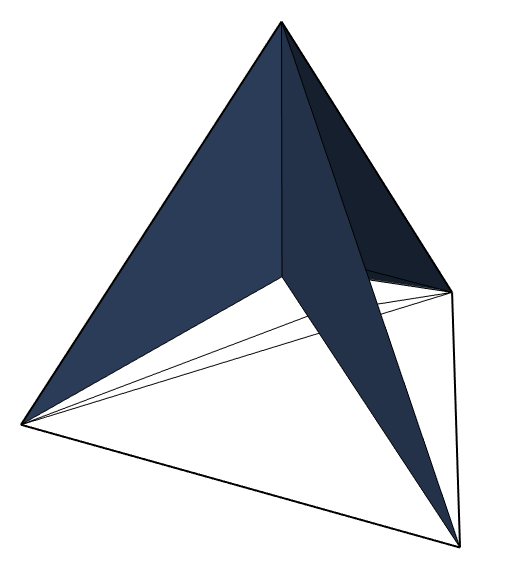}
  }%
  \hspace{25pt}
  \subfigure[][]{%
  \label{fig:triangulation_sigma_t_Tab_basso}%
  \includegraphics[scale=0.16]{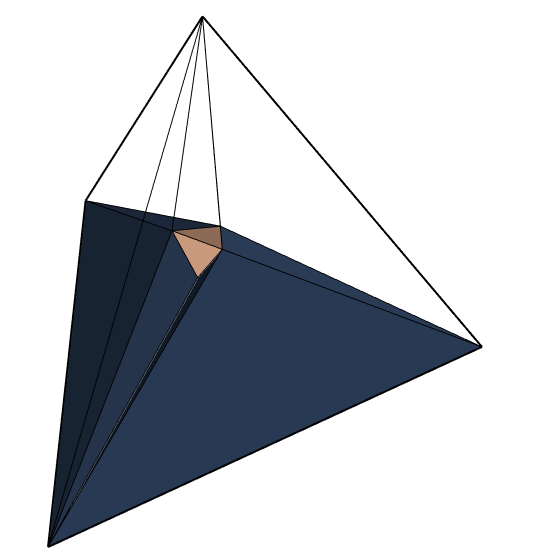}
  }%
  \hspace{19pt}
  \subfigure[][]{%
  \label{fig:triangulation_sigma_t_misto}%
   \includegraphics[scale=0.15]{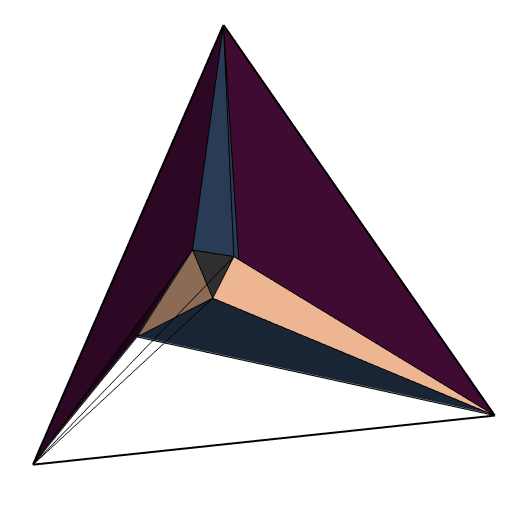}
   }\\
\caption{The triangulation of $\Sigma^{t}$ (not $\Sigma$) seen as a topological three-dimensional object. Each subfigure has different elements highlighted: in \subref{fig:triangulation_sigma_t_plain} the triangles $L^-_a$ are highlighted;
in \subref{fig:triangulation_sigma_t_T-} the four tetrahedra $N^{t-}_{a}$ are highlighted;
 in \subref{fig:triangulation_sigma_t_T+} three tetrahedra $N^{t+}_{a}$ out of four are highlighted; 
 in \subref{fig:triangulation_sigma_t_Tab_alto} three tetrahedra $N^t_{ab}$ out of six are highlighted;
 in \subref{fig:triangulation_sigma_t_Tab_basso} the remaining three tetrahedra $N^t_{ab}$ are highlighted;
 in \subref{fig:triangulation_sigma_t_misto} two tetrahedra $N^{t-}_{a}$, two tetrahedra $N^{t+}_{a}$ and two tetrahedra $N^t_{ab}$ are highlighted.
  }%
\label{fig:triangulation_sigma_t}%
\end{figure}

The triangulation is then defined by the points $p_a^\epsilon$, the segments $s^\epsilon_{ab}$ and $s^t_{a,b}$, the triangles $L^\epsilon_{a}$ and $L^{t\epsilon}_{a,b}$, the tetrahedra $N^{t\epsilon}_{a}$ and $N^t_{ab}$, and their boundary relations:
\begin{subequations}\label{eq:def_triangulation}
\begin{align}
&\partial s^\epsilon_{ab}=(p^\epsilon_a,p^\epsilon_b), \\
&\partial s^t_{a,b}=(p^-_a,p^+_b)^{t};\\
&\partial L^\epsilon_{a}=(s^\epsilon_{bc},s^\epsilon_{cd},s^\epsilon_{db}),\\
&\partial L^{t+}_{a,b}=(s^+_{cd},s^t_{a,c},s^t_{a,d}), \\
&\partial L^{t-}_{a,b}=(s^-_{cd},s^t_{c,a},s^t_{d,a}); \\
&\partial N^{t\epsilon}_{a}=(L^{t\epsilon}_{a,b},L^{t\epsilon}_{a,c},L^{t\epsilon}_{a,d},L^{\epsilon}_{a}), \\
&\partial N^t_{ab}=(L^{t+}_{a,b},L^{t+}_{b,a},L^{t-}_{c,d},L^{t-}_{d,c}). 
\end{align}
\end{subequations}
The indices $a,b,c,d,t$ and $\epsilon$ follow the rules discussed in \cref{section:discretization}. Partial graphical representations of this triangulation can be found in \cref{fig:triangulation_Sigma_fancy,fig:triangulation_sigma_t}.

The graph $\Gamma$ dual to the triangulation of $\Sigma$, which is formed by nodes (dual to the tetrahedra of the triangulation) connected by links (dual to the triangles of the triangulation), is represented in \cref{fig:dual_sigma}. Using the same label to denote dual objects, it is straightforward to check that the notation in \cref{eq:def_triangulation} is consistent with the notation given in \cref{section:discretization} for $\Gamma$.

The cellular decomposition of $B$ can then be obtained defining internal two-dimensional surfaces $f_{a,b}$, internal three-dimensional cells $e^\epsilon_{a,b}$, internal four-dimensional cells $v^\epsilon_a$ and $v_{ab}$, and their boundary relations:
\begin{subequations}\label{eq:def_cellular_decomposition}
\begin{align}
&\partial f_{a,b}=(s^p_{a,b},s^f_{a,b}); \\
&\partial e^\epsilon_{a,b}=(L^{p\epsilon}_{a,b},L^{f\epsilon}_{a,b},f_{a,c},f_{a,d});\\
&\partial v^\epsilon_a=(N_a^{p\epsilon},N_a^{f\epsilon},e^\epsilon_{a,b},e^\epsilon_{a,c},e^\epsilon_{a,d}),\\
&\partial v_{ab}=(N^p_{ab},N^f_{ab},e^+_{a,b},e^+_{b,a},e^-_{c,d},e^-_{d,c}).
\end{align}
\end{subequations}
Notice that the internal two-dimensional surfaces are not triangles, that the internal three-dimensional cells are not tetrahedra and that the internal four-dimensional cells are not four-simplices. The cellular decomposition of $B$ is thus not a four-dimensional triangulation.

The two-complex $\mathcal{C}$, whose combinatorial definition is given in \cref{section:discretization}, can be equivalently defined (apart from its orientation) as the dual to the cellular decomposition of $B$ specified by \cref{eq:def_cellular_decomposition,eq:def_triangulation}.

\section{Graphical representation of \texorpdfstring{$\Gamma$}{\textGamma} and \texorpdfstring{$\mathcal C$}{C}}
\label{section:appendix_2}

\noindent
The combinatorial characterization of the two-complex given above is complete and compact, but it is difficult to visualize. We give here some graphical representations to help the geometrical intuition.

\begin{figure}
\begin{overpic}[width = 0.8 \columnwidth]{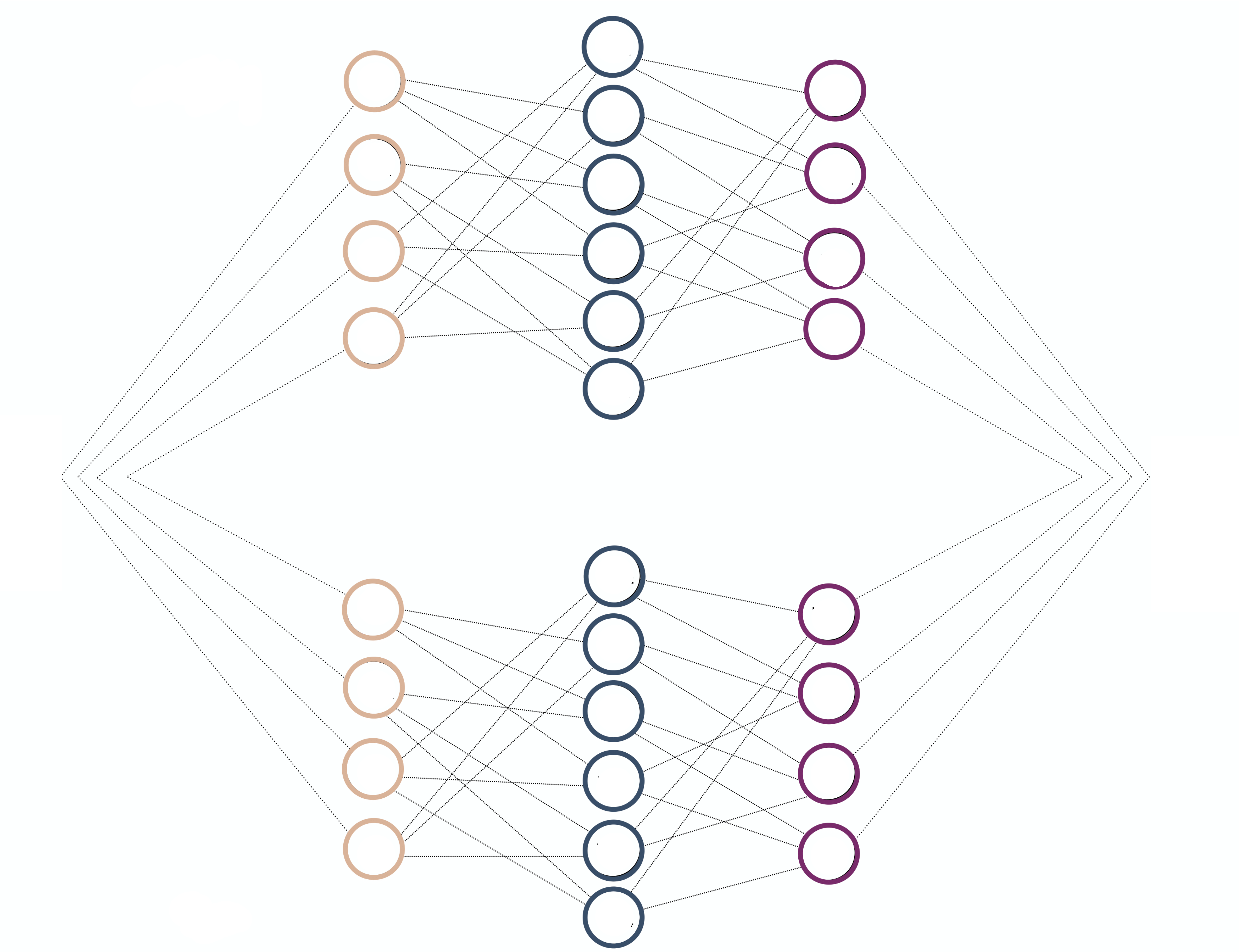}
    \put (65,0) {$\mathrm{N}_a^{p+}$}
    \put (27,0) {$\mathrm{N}_a^{p-}$}
    \put (65,75) {$\mathrm{N}_a^{f+}$}
    \put (27,75) {$\mathrm{N}_a^{f-}$}
    \put (47,-4) {$\mathrm{N}_{ab}^{p}$}
    \put (47,78) {$\mathrm{N}_{ab}^{f}$}
    \put (75,50) {$\mathrm{L}_a^{+}$}
    \put (16,50) {$\mathrm{L}_a^{-}$}
    \put (55,31) {$\mathrm{L}_{a,b}^{p +}$}
    \put (35,31) {$\mathrm{L}_{a,b}^{p -}$}
    \put (55,44) {$\mathrm{L}_{a,b}^{f +}$}
    \put (35,44) {$\mathrm{L}_{a,b}^{f -}$}
\end{overpic}
\caption{Two-dimensional representation of the graph $\Gamma$ dual to the triangulation of $\Sigma$.}
\label{fig:Gamma_2}      
\end{figure}

The graph $\Gamma$ dual to the triangulation of $\Sigma$ is depicted in \cref{fig:Gamma_2}. The lower part of the graph is on $\Sigma^p$, the upper is on $\Sigma^f$. The left part of the graph, with the four brown nodes $N^{t-}_a$, is the interior of the black hole; the right part of the graph, with the four red nodes $N^{t+}_a$, is the exterior of the black hole.  The six blue nodes $N^t_{ab}$ are intermediate (they represent tetrahedra cut by the horizon). 

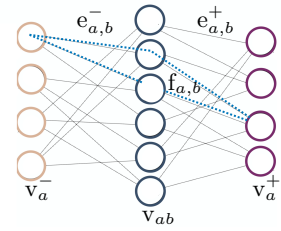
\begin{figure}
\begin{overpic}[width = 0.5 \columnwidth]{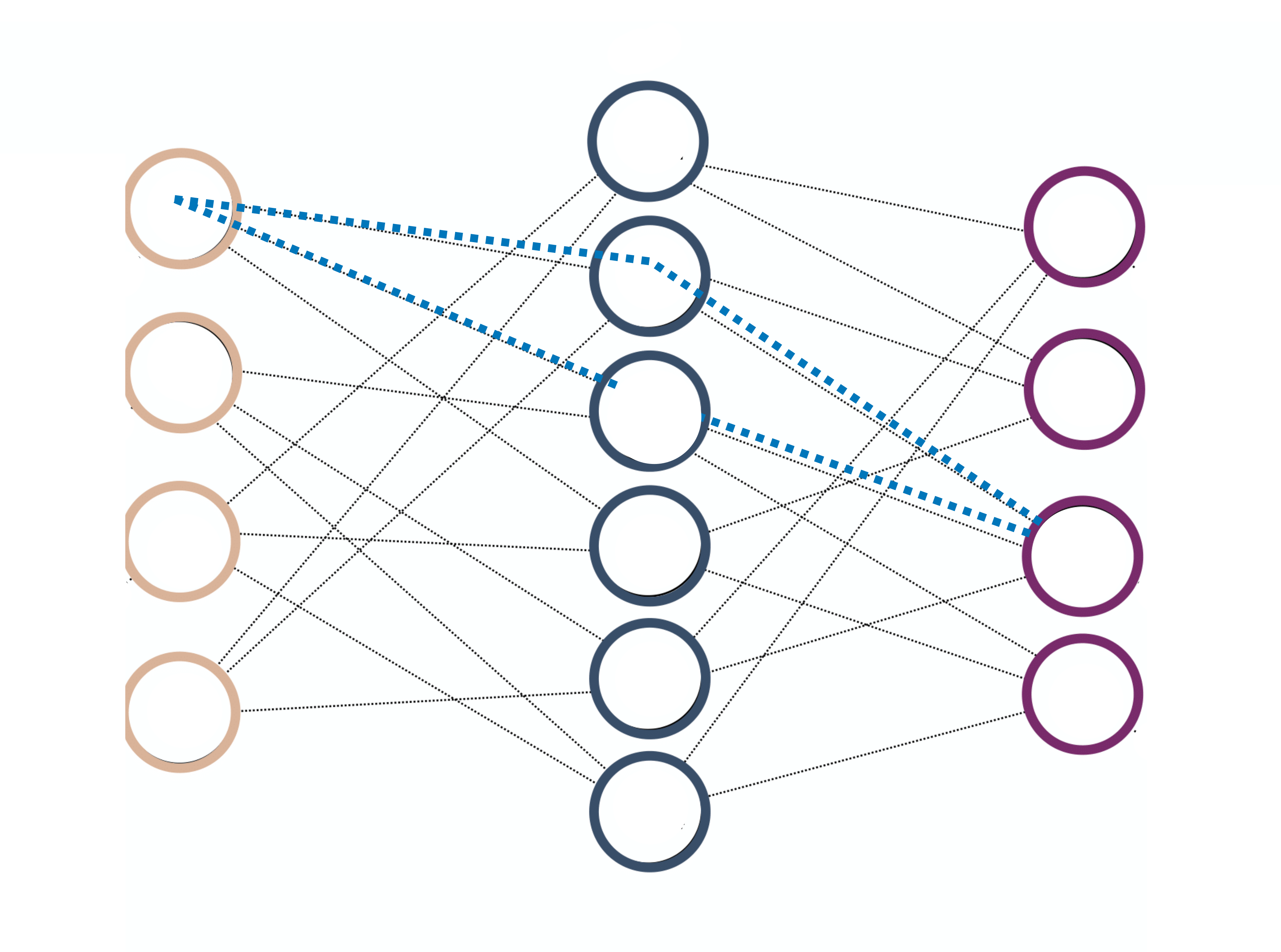}
    \put (82,9.5) {$\mathrm{v}^+_a$}
    \put (12,9.5) {$\mathrm{v}^-_a$}
    \put (48,1.5) {$\mathrm{v}_{ab}$}
    \put (65,61.5) {$\mathrm{e}^+_{a,b}$}
    \put (28,61.5) {$\mathrm{e}^-_{a,b}$}
    \put (56,42) {$\mathrm{f}_{a,b}$}
\end{overpic}
\caption{Two-dimensional representation of the internal component of the two-complex.}
\label{fig:internal_component_2}      
\end{figure}

The structure of the vertices and internal edges of the two-complex reproduces the structure of the past (or future) part of the graph $\Gamma$. This is depicted in \cref{fig:internal_component_2}, using for the vertices the same colour codes used for the nodes.
\begin{figure}
    \centering
    \includegraphics[scale=0.26]{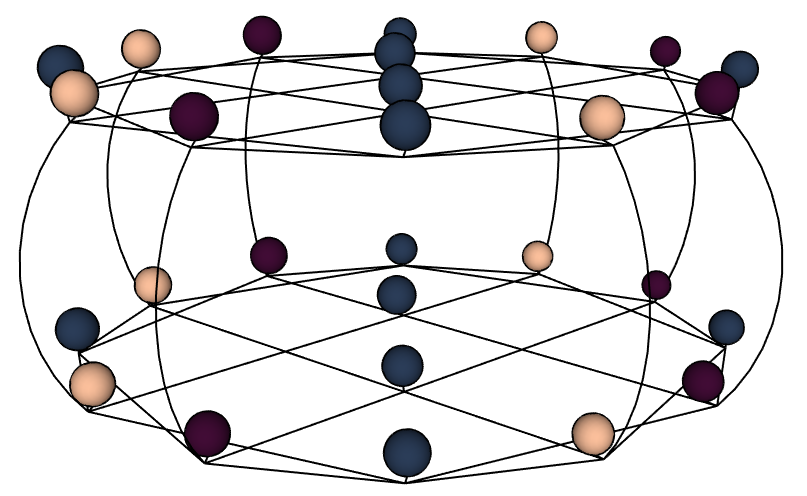}
    \caption{Three-dimensional representation of the graph $\Gamma$ dual to the triangulation of $\Sigma$; each node is represented as a sphere colored consistently with its dual tetrahedron in \cref{fig:triangulation_sigma_t}.}
    \label{fig:dual_sigma}
\end{figure}
The boundary edges are all "vertical": they connect the vertices in \cref{fig:internal_component_2} with the corresponding past and future nodes in \cref{fig:Gamma_2}. 
All the internal faces are "horizontal": one of them is depicted in figure \eqref{fig:internal_component_2}. All the boundary faces are "vertical" and they are of course in one-to-one correspondence with the links of the graph $\Gamma$.

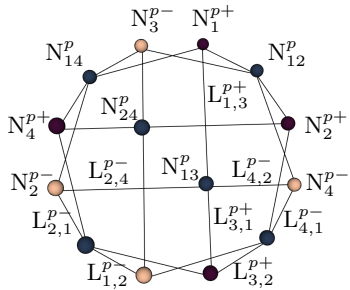
\begin{figure}
    \centering
    \begin{overpic}[scale=0.24]{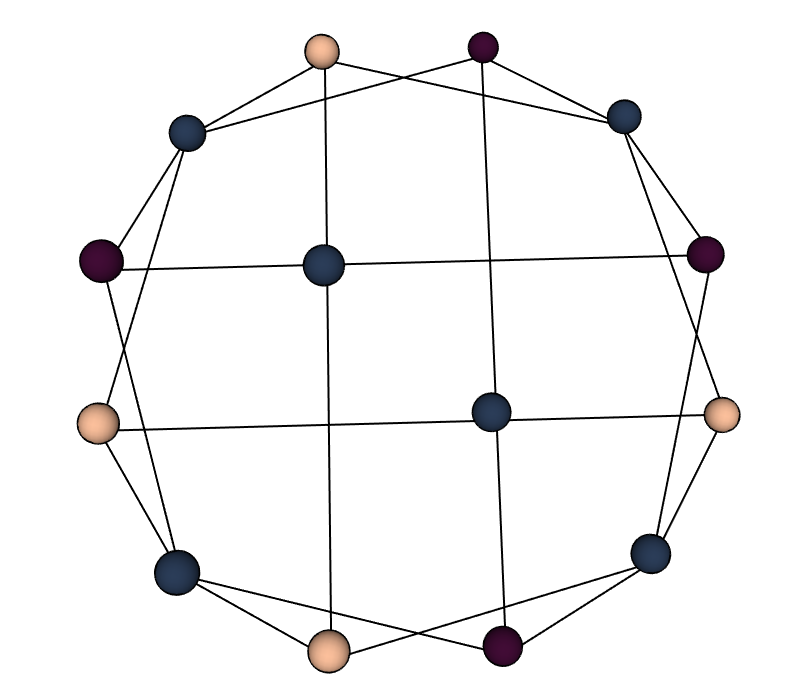}
    \put (57,87) {$\mathrm{N}^{p+}_1$}
    \put (37,87) {$\mathrm{N}^{p-}_3$}
    \put (94,52) {$\mathrm{N}^{p+}_2$}
    \put (94.5,33) {$\mathrm{N}^{p-}_4$}
    \put (-3.5,52) {$\mathrm{N}^{p+}_4$}
    \put (-2.5,33) {$\mathrm{N}^{p-}_2$}
    \put (9,76) {$\mathrm{N}^{p}_{14}$}
    \put (82,76) {$\mathrm{N}^{p}_{12}$}
    \put (27,58) {$\mathrm{N}^{p}_{24}$}
    \put (47,38.5) {$\mathrm{N}^{p}_{13}$}
    \put (4.5,20.5) {$\mathrm{L}^{p-}_{2,1}$}
    \put (22.5,3.5) {$\mathrm{L}^{p-}_{1,2}$}
    \put (87,21) {$\mathrm{L}^{p-}_{4,1}$}
    \put (70.5,4) {$\mathrm{L}^{p+}_{3,2}$}
    \put (23,37) {$\mathrm{L}^{p-}_{2,4}$}
    \put (70,37.5) {$\mathrm{L}^{p-}_{4,2}$}
    \put (64,22) {$\mathrm{L}^{p+}_{3,1}$}
    \put (62,62) {$\mathrm{L}^{p+}_{1,3}$}
    \end{overpic}
    \caption{Two-dimensional representation of the past component of the graph $\Gamma$ dual to the triangulation of $\Sigma$.}
    \label{fig:dual_sigma_t_2D}
\end{figure}

To visualize the faces, it is more convenient to shift to a three-dimensional representation and give up the radial ordering from interior to exterior. The graph $\Gamma$ can be then represented as in \cref{fig:dual_sigma}. The upper part of the picture still contains the future objects and the lower part the past objects. The blue dots represent the nodes $\mathrm{N}^{t}_{ab}$, the brown dots represent the nodes $\mathrm{N}^{t-}_a$ and the red dots represent the nodes $\mathrm{N}^{t+}_a$, as in \cref{fig:Gamma_2}, but the interior-middle-exterior order is not respected.  The links $\mathrm{L}^{\epsilon}_{a}$ vertically connect the nodes $\mathrm{N}^{p\epsilon}_a$ in the past to the nodes $\mathrm{N}^{f\epsilon}_a$ in the future. The pattern of the links $\mathrm{L}^{t\epsilon}_{a,b}$, which is the same on both the past and the future components of the graph, can be better appreciated restricting only to the past (or future) component depicted in \cref{fig:dual_sigma_t_2D}.

\begin{figure}[b]
  \includegraphics[keepaspectratio=true,scale=0.28]{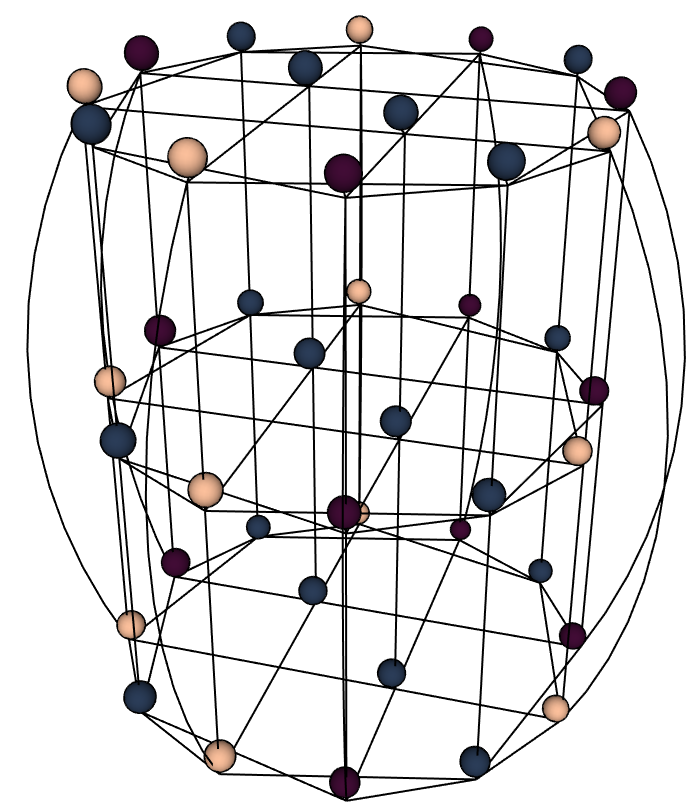}
\caption{Three-dimensional representation of the one-skeleton of the two-complex $\mathcal{C}$; notice that although the internal vertices and the boundary nodes are graphically depicted in the same way, they are two very distinct objects (the same applies also to edges and links).}
\label{fig:twocomplex}        
\end{figure}

The one-skeleton of the two-complex is represented in \cref{fig:twocomplex}.   
Careful: in \cref{fig:twocomplex} the dots of the upper and lower layers are nodes, while those of the intermediate layer are vertices. The past (lower) and the future (upper) layers are in fact the past and the future components of the boundary graph $\Gamma$ in \cref{fig:dual_sigma}. The intermediate layer represents the vertices and the internal edges of the two-complex.  

\begin{figure}[t]
  \includegraphics[scale=0.2]{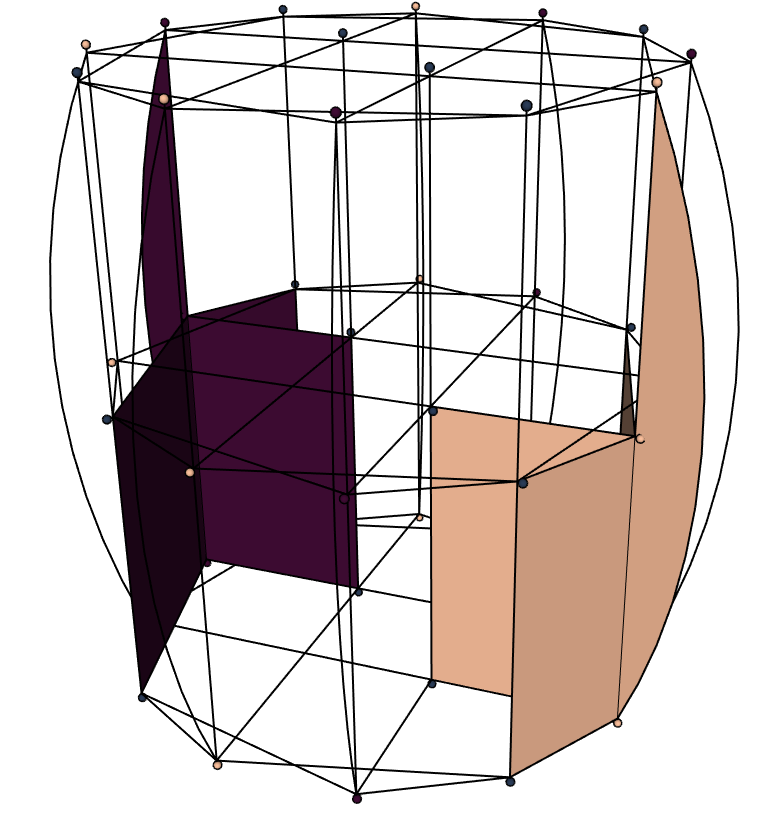}
\captionof{figure}{Graphical representation of the boundary faces (in red) $\mathrm{F}^{p+}_{4,1}$, $\mathrm{F}^{p+}_{4,2}$, $\mathrm{F}^{p+}_{4,3}$, $\mathrm{F}^+_{4}$ and (in brown) $\mathrm{F}^{p-}_{4,1}$, $\mathrm{F}^{p-}_{4,2}$, $\mathrm{F}^{p-}_{4,3}$, $\mathrm{F}^-_{4}$.}
\label{fig:spinfoam_boundary_faces}      
\end{figure}

\begin{figure}[b]
\centering
  \subfigure[][]{%
  \label{fig:internal_faces_1}%
  \includegraphics[scale=0.2]{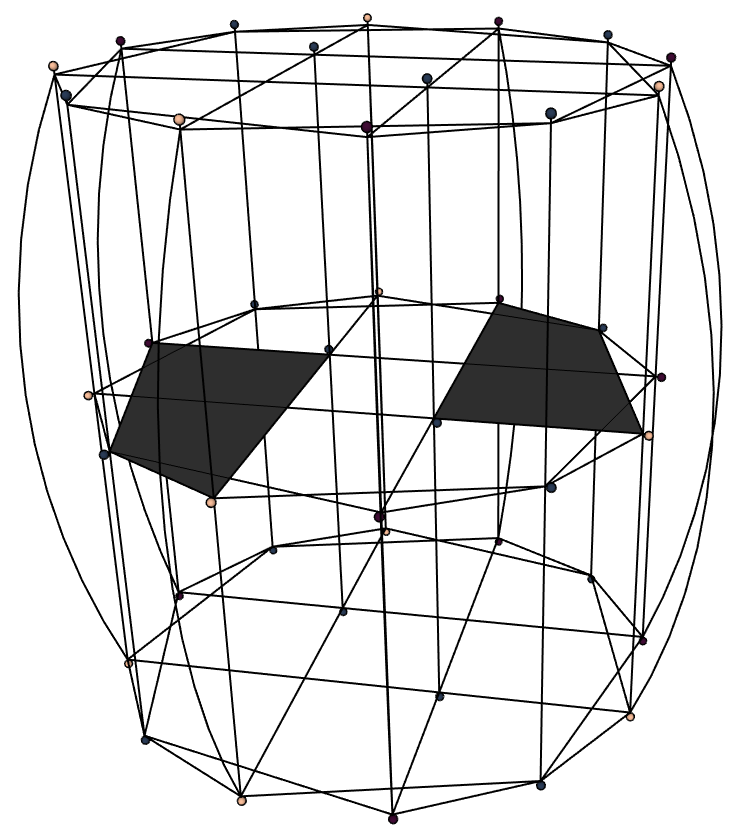}
  }%
  \hspace{20pt}
  \subfigure[][]{%
  \label{fig:internal_faces_2}%
  \includegraphics[scale=0.2]{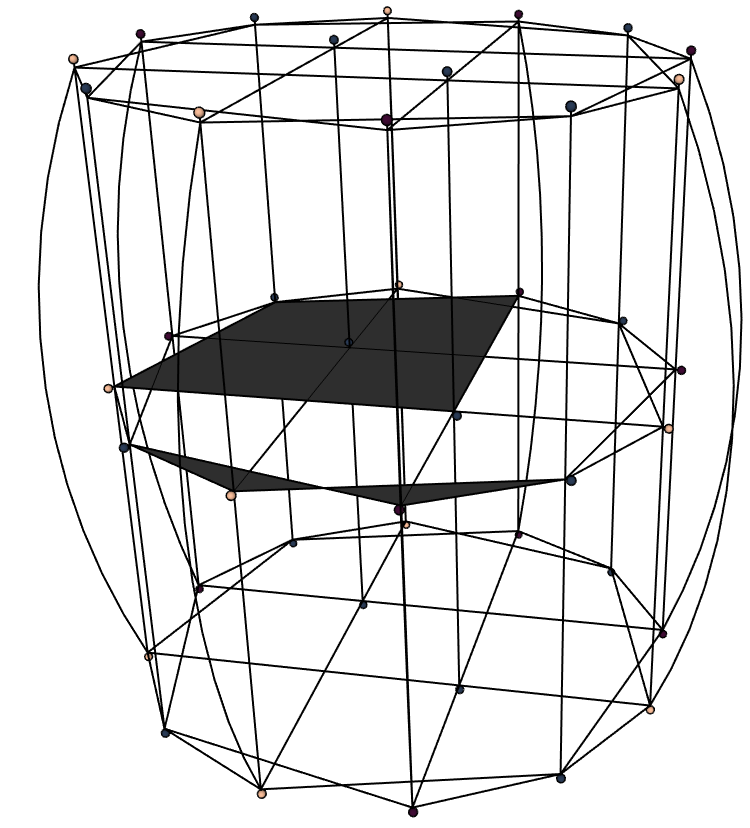}
  }%
\caption{Graphical representation of the internal faces $\mathrm{f}_{1,4}$, $\mathrm{f}_{4,1}$ (in \subref{fig:internal_faces_1}) and $\mathrm{f}_{1,2}$, $\mathrm{f}_{3,1}$ (in \subref{fig:internal_faces_2}).}
\label{fig:internal_faces}  
\end{figure}

The boundary edges $\mathrm{E}^{t\epsilon}_a$ and $\mathrm{E}^t_{ab}$ can be recognized in \cref{fig:twocomplex} as the edges connecting the internal component of the two-complex to both the past and the future components of the boundary. This construction completely specifies the one-skeleton of the two-complex $\mathcal{C}$.

The graphical representation of the boundary faces $\mathrm{F}^{\epsilon}_{a}$ and $\mathrm{F}^{t\epsilon}_{a,b}$ is easily obtained using their definition given in \cref{section:discretization}. Some of them are depicted in \cref{fig:spinfoam_boundary_faces}. 

The internal faces, although slightly more difficult to represent, can be found in the same way. Some of them are reported in \cref{fig:internal_faces}. The strange nature of their graphical representation (some of them intersect each other and some of them have strange shapes) is just a consequence of the fact that we are representing a four-dimensional object in three dimensions and it has no physical meaning.

\section{Transition amplitudes in covariant loop quantum gravity}
\label{section:appendix_3}

\noindent
We briefly review how transition amplitudes are expressed in covariant loop quantum gravity focusing on the EPRL-KKL transition amplitudes. For more details, we refer the reader to~\cite{book:Rovelli_Vidotto_CLQG,Perez:2012wv,article:Kaminski_kisielowski_Lewandowski_2010_generalized_spinfoam,article:Ding_Han_Rovelli_2011_generalized_spinfoams}.

Given a compact region of spacetime $B$ with boundary $\Sigma=\partial B$ and an arbitrary oriented graph $\Gamma \in \Sigma$, the boundary Hilbert space of a truncation of the complete quantum theory that consider only the degrees of freedom coming from $\Gamma$ is $H_{\Gamma}=L^{2} \left[ \mathrm{SU}(2)^{L}/\mathrm{SU}(2)^{N} \right]_{\Gamma}$, where $L$ and $N$ are respectively the total number of links and nodes in $\Gamma$. A boundary state, i.e. an element of the boundary space, is a square integrable function $\psi(h_{\ell})$ that is gauge invariant at every node $\mathrm{n} \in \Gamma$. Each $h_{\ell}\in\mathrm{SU}(2)$ can be seen as the holonomy of the Ashtekar-Barbero connection between two nodes of $\Gamma$. 

A transition amplitude associated to each state in the boundary Hilbert space can then be given in terms of an arbitrary oriented two-complex $C$ whose boundary graph is $\Gamma$. Let $\mathrm f$, $\mathrm e$, $\mathrm v \in C$ denote respectively a face, an edge and a vertex of $C$. To each internal edge $\mathrm{e}$ linking two vertices $\mathrm{v}$ and $\mathrm{v'}$ are assigned two $\mathrm{SL}(2,\mathbb{C})$ elements $g_{ve}=g^{-1}_{ev}$ and $g_{\mathrm{e}\mathrm{v}'}=g^{-1}_{\mathrm{v}'\mathrm{e}}$. To each boundary edge $\mathrm{E}$, edges that link an internal vertex $\mathrm{v}$ and a node $\mathrm{n}$, is assigned one $\mathrm{SL}(2,\mathbb{C})$ element $g_{\mathrm{vn}}=g^{-1}_{\mathrm{nv}}$. 

An internal face is denoted as $\mathrm{f}\in B$ and a boundary face, a face that contains a link, is denoted as $\mathrm{F}\in \Gamma$. Given an arbitrary boundary state $\ket{\psi} \in H_{\Gamma}$, the theory associate to it the amplitude

\be
\bra{W_C}\ket{\psi}=\int_{\mathrm{SU}(2)} \diff  h_{\ell} \: W_C ( h_{\ell})\, \psi( h_{\ell})\:\: ,
\label{equation:appendix_LQG_state_amplitude1}
\ee
where the two-complex amplitude $W_C ( h_{\ell} )$ is defined as (neglecting the normalization constant)
\be
W_C ( h_{\ell}  )=  \int_{\mathrm{SU}(2)}  \diff h_{\mathrm{fv}} \displaystyle\prod_{\mathrm{f}\in C} \delta \Big( \displaystyle\prod_{\mathrm{v}\in \mathrm{f}} h_{\mathrm{fv}}\Big) \displaystyle\prod_{\mathrm{v}\in C} A_{\mathrm{v}} ( h_{\mathrm{fv}} )\: .
\label{equation:appendix_LQG_2complex_amplitude_vertex_style}
\ee
The vertex amplitude $A_{\mathrm{v}} ( h_{\mathrm{fv}} )$ is given by
\be
\begin{split}
A_{\mathrm{v}} ( h_{\mathrm{fv}} )= &\displaystyle\sum_{j_{\mathrm{f}}}  \int_{\mathrm{SL}(2,\mathbb{C})} \displaystyle\diff g'_{\mathrm{ve}} \displaystyle\prod_{\mathrm{f}\ni \mathrm{v}} \Big[ d_{j_\mathrm{f}} \\
&\times \Tr\Big(\Df(g_{\mathrm{e'}_{\mathrm{f}} \mathrm{v}}g_{\mathrm{v} \mathrm{e}_{\mathrm{f}}}) D^{(j_{\mathrm{f}})}(h_{\mathrm{fv}}) \Big) \Big]\: .
\label{equation:appendix_LQG_vertex_amplitude_0}
\end{split}
\ee
The prime on $\diff g'_{\mathrm{ve}}$ means that, fixing $\mathrm{v}$, the integration is over all possible $g_{\mathrm{ev}}$ except one. The edges $\mathrm{e}_{\mathrm{f}}$ and $\mathrm{e'}_{\mathrm{f}}$ are the two edges in $\mathrm{f}$ that have the vertex $\mathrm{v}$ as, respectively, target and source. The matrix $D^{(j)}$ is the Wigner matrix of the $d_j$-dimensional ($d_j=2j+1$) representation of $\mathrm{SU}(2)$ acting on $H_j$. The matrix $\D$ is the $d_j \times d_j$ matrix 
\be
\quad\quad(\D(g))_{mn} \equiv {\cal D}^{(\gamma j, j)}_{jm\: jn} (g)\, , \quad\quad  g\in \mathrm{SL}(2,\mathbb{C}) \, ,
\label{calD2}
\ee
where ${\cal D}^{(p, k)}_{jm\: j'n}$ are the matrix elements of the $(p,k)$ unitary representation of the principal series of $\mathrm{SL}(2,\mathbb{C})$ in the canonical basis that diagonalize the operators $L^2$ and $L_z$ of the $\mathrm{SU}(2)$ subgroup. 
Finally, $\gamma$ is the Barbero-Immirzi parameter.

This is the two-complex amplitude expressed in terms of elementary vertex amplitudes. Inserting \cref{equation:appendix_LQG_vertex_amplitude_0} in \cref{equation:appendix_LQG_2complex_amplitude_vertex_style} and performing the integrations over $h_{\mathrm{fv}}$, it is possible to express the two-complex amplitude in terms of elementary face amplitudes. The result is

\be
\begin{split}
W_C (  h_{\ell} )=  \int_{\mathrm{SL}(2,\mathbb{C})} & \displaystyle\prod_{\mathrm{v}\in C} \diff g'_{\mathrm{ve}}  \displaystyle\prod_{\mathrm{f}\in B} A_{\mathrm{f}}  (g_{\mathrm{ve}} ) \\ & \times \prod_{\mathrm{F}\in \Gamma} A_{\mathrm{F}} ( h_{\ell_{\mathrm{F}}} , g_{\mathrm{ve}} )\: ,
\label{equation:appendix_LQG_2complex_amplitude_face_style}
\end{split}
\ee
where the amplitude of an internal face is 
\be
\begin{split}
A_{\mathrm{f}} & (g_{\mathrm{ve}} )= \displaystyle\sum_{j_{\mathrm{f}}} d_{j_{\mathrm{f}}}  \Tr\Big[ \Df (g_{\mathrm{ev}}g_{\mathrm{ve}'}) 
\\  & \times
 \Df (g_{\mathrm{e}'\mathrm{v}'}g_{\mathrm{v}'\mathrm{e}''}) 
 \cdots \Df
(g_{\mathrm{e}^{(n)}\mathrm{v}^{(n)}}g_{\mathrm{v}^{(n)}\mathrm{e}})  \Big]\: ,
\end{split}
\label{equation:appendix_LQG_internal_face_amplitude}
\ee
and the amplitude of a face with one link in the boundary is 
\be
\begin{split}
A_{\mathrm{F}}(h_{\ell_{\mathrm{F}}} & , g_{\mathrm{ve}} )=   \displaystyle\sum_{j_{\mathrm{F}}} d_{j_{\mathrm{F}}} \Tr\Big[ \DF
(g_{\mathrm{n}_t \mathrm{v}}g_{\mathrm{ve}'})  \\
\times & \DF
(g_{\mathrm{e}'\mathrm{v}'}g_{\mathrm{v}'\mathrm{e}''})
 \cdots \DF
(g_{\mathrm{e}^{(n)}\mathrm{v}^{(n)}}g_{\mathrm{v}^{(n)}\mathrm{n}_s}) \\
\times &
D^{(j_{\mathrm{F}})} (h_{\ell_{\mathrm{F}}})  \Big]\: .
\end{split}
\label{equation:appendix_LQG_boundary_face_amplitude}
\ee
The quantities $\mathrm{n}_s$ and $\mathrm{n}_t$ represent the nodes that are, respectively, source and target of the link $\ell$. The value of the label $(n)$ in $\mathrm{e}^{(n)}$ and $\mathrm{v}^{(n)}$ is fixed for each face by the topology of $C$ and $\Gamma$.

A third way to express the two-complex transition amplitude is the coherent representation in terms of $\mathrm{SU}(2)$ coherent states $\ket{j,\vec{n}} \in H_j$. Starting from either \cref{equation:appendix_LQG_2complex_amplitude_vertex_style} or \cref{equation:appendix_LQG_2complex_amplitude_face_style}, the two-complex transition amplitude can be expressed in its coherent representation by explicitly performing the traces in terms of coherent states and inserting coherent resolutions of the identity on $H_j$,
\be
\label{eq:resolution_identity_coherent_states}
\mathbbm{1}_j = \frac{d_j}{4\pi} \int_{S^2} \diff^2 \vec{n}
\:\ket{j,\vec{n}}\bra{j,\vec{n}}\, ,
\ee
between all the matrices. Rearranging the result in terms of elementary vertex amplitudes the coherent two-complex transition amplitude can be written as 
\be
\begin{split}
W_C (  h_{\ell} )= &\sum_{j_{\mathrm{f}}} \prod_{\mathrm{f}\in C} d_{j_{\mathrm{f}}} \prod_{\mathrm{e}\in C} \prod_{\mathrm{f}\ni \mathrm{e}} \bigg[  \int_{S^2} 
\diff^2 \vec{n}_{\mathrm{ef}} \:\frac{d_{j_{\mathrm{f}}}}{4\pi} \bigg]\\
\times & \prod_{\ell\in \Gamma} \bra{j_{\mathrm{f}_{\ell}},\vec{n}_{\mathrm{E}_{\mathrm{n}_s} \mathrm{f}_{\ell}}}
D^{(j_{\mathrm{f}_{\ell}})}( h_{\ell})
\ket {j_{\mathrm{f}_{\ell}},\vec{n}_{\mathrm{E}_{\mathrm{n}_t} \mathrm{f}_{\ell}}}     \\
\times & \prod_{\mathrm{v}\in C} 
A_{\mathrm{v}} \big( j_{\mathrm{f}} , \vec{n}_{\mathrm{ef}} \big)
\: ,
\end{split}
\label{equation:appendix_LQG_2complex_amplitude_coherent_style}
\ee
where the coherent vertex amplitude is
\be
\begin{split}
A_{\mathrm{v}}  & \big( j_{\mathrm{f}} , \vec{n}_{\mathrm{ef}} \big) =  \int_{\mathrm{SL}(2,\mathbb{C})} \displaystyle\diff g'_{\mathrm{ve}}\\
\times & \displaystyle \prod_{\mathrm{f}\ni \mathrm{v}}
\bra{j_{\mathrm{f}},\vec{n}_{\mathrm{e}_{\mathrm{f}} \mathrm{f}}}
\Df(g_{\mathrm{e}_{\mathrm{f}} \mathrm{v}}g_{\mathrm{v} \mathrm{e'}_{\mathrm{f}}})
\ket{j_{\mathrm{f}},\vec{n}_{\mathrm{e'}_{\mathrm{f}} \mathrm{f}}}
\: .
\label{equation:appendix_LQG_vertex_amplitude}
\end{split}
\ee
The label $\mathrm{E}_{\mathrm{n}}$ is used to denote the boundary edge that has the node $\mathrm{n}$ in its boundary.

A priori, this construction assigns to each half-edge $\mathrm{ev}$ (assuming the same orientation for each face $\mathrm{f}$ having $\mathrm{e}$ in its boundary) the state
\be
\bigotimes_{\mathrm{f}\ni  \mathrm{e}} \ket{j_{\mathrm f}, \vec{n}_{\mathrm{ef}}}\in \bigotimes_{\mathrm{f}\ni  \mathrm{e}} H_{j_{\mathrm f}}\, .
\ee
However, due to $\mathrm{SU}(2)$ invariance induced on the half-edge $\mathrm{ev}$ by the integration over $g_{\mathrm{ev}}\in\mathrm{SL}(2,\mathbb{C})$, the state assigned to $\mathrm{ev}$ can be seen as a coherent intertwiner \cite{article:Livine_Speziale_2007_new_spinfoam_vertex} 
\be
\int_{\mathrm{SU}(2)} \diff h \bigotimes_{\mathrm{f}\ni  \mathrm{e}} h \triangleright \ket{j_{\mathrm f}, \vec{n}_{\mathrm{ef}}}
\in 
\mathrm{Inv}_{\mathrm{SU}(2)} \Big[ \bigotimes_{\mathrm{f}\ni  \mathrm{e}} H_{j_{\mathrm f}} \Big]\, .
\ee
This suggests a fourth way to express the transition amplitude: instead of using the over-complete basis of the coherent intertwiners we could use a basis of intertwiners
\be
\ket{i_{\mathrm e}}
\in 
\mathrm{Inv}_{\mathrm{SU}(2)} \Big[ \bigotimes_{\mathrm{f}\ni  \mathrm{e}} H_{j_{\mathrm f}} \Big]
\ee
giving the following resolution of the identity
\be
\mathbbm{1}= \sum_{i_{\mathrm e}} 
\ket{i_{\mathrm e}}\bra{i_{\mathrm e}}\, .
\ee
The two-complex transition amplitude can be then schematically rewritten as
\be
\begin{split}
W_C (  h_{\ell} )= &\sum_{j_{\mathrm{f}}, i_{\mathrm e}} \prod_{\mathrm{f}\in C} d_{j_{\mathrm{f}}} 
\bra{\prod_{\mathrm E \in C}i_{\mathrm{E}} \, }\ket{\:\prod_{\ell\in \Gamma} D^{(j_{\mathrm{f}_{\ell}})}( h_{\ell})}   
\\\times & 
\prod_{\mathrm{v}\in C} 
A_{\mathrm{v}} \big( j_{\mathrm{f}} , i_{\mathrm{e}} \big)
\: ,
\end{split}
\label{equation:appendix_LQG_2complex_amplitude_intertwiner_style}
\ee
where the spin-intertwiner vertex amplitude is
\be
\begin{split}
A_{\mathrm{v}}   \big( j_{\mathrm{f}} , i_{\mathrm{e}} \big) = &  \int_{\mathrm{SL}(2,\mathbb{C})} \displaystyle\diff g'_{\mathrm{ve}}\\
\times & \bra{\prod_{\mathrm e \ni \mathrm{v}} i_{\mathrm{e}} \, }\ket{\:\prod_{\mathrm{f}\ni \mathrm{v}} \Df(g_{\mathrm{e}_{\mathrm{f}} \mathrm{v}}g_{\mathrm{v} \mathrm{e'}_{\mathrm{f}}})}  
\label{equation:appendix_LQG_vertex_amplitude_intertwiner}
\end{split}
\ee
and the bra-ket notation indicates the index contraction (between the intertwiners and the matrix elements) dictated by the topology of the two-complex.

\bibliographystyle{utcaps}
\bibliography{references.bib}

\end{document}